\documentclass[aps,pra,showpacs,amsmath,amssymb,amsfonts,lengthcheck,twocolumn,longbibliography]{revtex4-1}
\usepackage{dcolumn}    
\usepackage{bm}
\usepackage{subfigure}
\usepackage{float}
\usepackage{verbatim}
\usepackage{graphicx}
\usepackage{amsmath}
\usepackage{stmaryrd}
\usepackage{amssymb}
\usepackage{epsf}
\usepackage{dsfont}
\usepackage{color}
\usepackage{xcolor}
\usepackage[colorlinks=true,linkcolor=blue,urlcolor=blue,citecolor=blue,pdfusetitle]{hyperref}
\usepackage[normalem]{ulem}

\usepackage{fmtcount}
\usepackage{cleveref}

\makeatletter
\newcommand\footnoteref[1]{\protected@xdef\@thefnmark{\ref{#1}}\@footnotemark}
\makeatother
\newcommand\myeqref[1]{
	Eq. (\textup{\ref{#1}})
}

\newcommand{\blah}{blah\\blah\\blah\\blah\\blah.}

\newcommand{\aref}[1]{{Appendix~\hyperref[#1]{A}}}
\newcommand{\bref}[1]{{Appendix~\hyperref[#1]{B}}}
\newcommand{\la}{\left\langle}
\newcommand{\ra}{\right\rangle}
\newcommand{\bra}[1]{\left\langle #1\right|}
\newcommand{\ket}[1]{\left|#1\right\rangle}

\newcommand{\tr}[1]{\mathrm{tr}\left\{#1\right\}}

\newcommand{\pd}{\partial}

\newcommand{\etal}{\textit{et al. }}
\newcommand{\e}[1]{\exp{\left(#1\right)}}
\newcommand{\lo}[1]{\ln{\left(#1\right)}}

\newcommand{\com}[2]{\left[#1,\,#2\right]}

\newcommand{\bla}{bla\\bla\\bla\\bla\\bla}
\newcommand{\ex}[1]{\exp{\left(#1\right)}}

\newcommand{\PRA}{Phys. Rev. A }
\newcommand{\PRB}{Phys. Rev. B }
\newcommand{\PRD}{Phys. Rev. D }
\newcommand{\PRE}{Phys. Rev. E }
\newcommand{\PRL}{Phys. Rev. Lett. }

\newcommand{\RMP}{Rev. Mod. Phys. }
\newcommand{\NJP}{New. J. Phys. }

\newcommand{\mc}[1]{\mathcal{#1}}
\newcommand{\mbb}[1]{\mathbb{#1}}

\newcommand{\mrm}[1]{\mathrm{#1}}

\begin{document}
\title{Information scrambling vs. decoherence -- two competing sinks for entropy}

\author{Akram Touil}
\email{akramt1@umbc.edu}
\author{Sebastian Deffner}
\email{deffner@umbc.edu}
\affiliation{Department of Physics, University of Maryland, Baltimore County, Baltimore, MD 21250, USA}

\begin{abstract}
A possible solution of the information paradox can be sought in quantum information scrambling. In this paradigm, it is postulated that all information entering a black hole is rapidly and chaotically distributed across the event horizon making it impossible to reconstruct the information by means of any local measurement. However, in this scenario the effects of decoherence are typically ignored, which may render information scrambling moot in cosmological settings. In this work, we develop key steps towards a thermodynamic description of information scrambling in open quantum systems. In particular, we separate the entropy production into contributions arising from scrambling and decoherence, for which we derive statements of the second law. This is complemented with a numerical study of the Sachdev-Ye-Kitaev, Maldacena-Qi, XXX, mixed field Ising, Lipkin-Meshkov-Glick models in the presence of decoherence in energy or computational basis.
\end{abstract}
\date{\today}
\maketitle

\section{Introduction}

Ever since their inception \cite{einstein1915,carroll2004introduction}, black holes have sparked the imagination of physicists, sci-fi authors, and artists alike. Thus, it was a particularly big event when only last year the first image of a black hole was released \cite{event2019first}, which gave a scientifically grounded  face to these mysterious regions of space. A major obstacle in the understanding of space-time singularities is that the usual paradigms of measurement cannot be applied directly. All measurements in physics rely either on intercepting some intrinsic emission, or on observing the response to an outside perturbation.  The only emission from a black hole is the rather faint Hawking radiation, which is very hard to observe \cite{Steinhauer2016}. The second technique is further debilitated by the fact that to date no clear consensus has emerged on what exactly happens to a signal sent towards a black hole once it crosses the event horizon \cite{paradox1,paradox2,preskill1992black}.

A possible resolution of this ``information paradox'' has been suggested by quantum information theory \cite{unit1,unit2,unit3}. It has been argued \cite{almheiri2020replica} that all information crossing the event horizon is essentially instantaneously and chaotically ``scrambled'' across the entirety of the horizon, and eventually leaves through wormholes connecting the interior of the black hole with the Hawking radiation. Whether or not this is what really happens in the cosmos remains to be seen. However, the concept of \emph{quantum information scrambling} has opened the door to all kinds of fundamental questions. Rather remarkably, its study has attracted significant attention in various other fields of physics, including but not limited to high energy physics \cite{Hosur2016,maldacena2016bound,hashimoto2017out}, quantum information theory \cite{swingle2016measuring,swingle2018unscrambling,Yoshida2019,touil2020quantum}, and condensed matter and quantum many body theory \cite{chen2016universal,Iyoda2018,seshadri2018tripartite,chen2018operator,Alba2019,Chenu2019workstatistics,bergamasco2019otoc,nakamura2019universal}.

Only very recently, the existence of quantum information scrambling was demonstrated in an experiment with ion traps \cite{landsman2019verified}. In this experiment, the scrambling of information was verified through a teleportation scheme, analogous to two black holes scrambling and exchanging information through an Einstein-Rosen (ER) bridge, in the context of the ``ER=EPR'' conjecture \cite{epr}. As one would expect when working with any real quantum system, Landsman \etal \cite{landsman2019verified} had to overcome the detrimental effects of decoherence. This is particularly noteworthy, once one realizes that quantum information scrambling is rooted in the spread of entanglement, which is notoriously hard to preserve in the presence of noise.  In actuality, similar considerations may be instrumental when analyzing scrambling in black holes. Indeed it has been shown that the intense gravitational field of black holes may act as a decoherence channel \cite{gravity1,gravity2,gravity3} in itself. Therefore, the natural question arises if and to what extent information scrambling is immune to the intricate effects of decoherence. 

Curiously, most of the work done in the literature focuses only on  information scrambling in closed systems, i.e., for unitary dynamics, using the out-of-time-ordered correlator (OTOC) as a quantifier. One of the very few analyses of open system dynamics was published in Ref.~\cite{prb}. Zhang \etal \cite{prb} reported a numerical study of the effects of dissipation on information scrambling in quantum chaotic systems. To this end, they used a specific measurement technique to compute the OTOC \cite{prbtechnique}. This allowed them to isolate dissipation from scrambling by computing a ratio of OTOCs (for a given set of operators). Reference~\cite{prb} found that while dissipation diminishes scrambling, there are domains in parameter space in which information is still distributed across all regions in Hilbert space.

In the present work, we tackle an even broader class of problems: information scrambling in general open quantum systems. Special emphasis is put on a quantum thermodynamics perspective, so that we can identify and separate the contributions of scrambling and decoherence to the irreversible entropy production. In a previous work \cite{touil2020quantum} we argued that the OTOC is a somewhat awkward quantity for thermodynamic analyses. Moreover, the OTOC alone is not enough to distinguish between unitary evolution, that induces scrambling, and the external noise, which is the origin of decoherence. The latter problem is of particular importance in experiments that were conducted to verify quantum information scrambling \cite{landsman2019verified,experiment}. In these experiments, additional quantities such as the teleportation fidelity \cite{landsman2019verified} and the R\'enyi-2 entropy \cite{experiment}, were computed to be able to distinguish scrambling from noise. However, these quantities are model specific and thus a more general, theoretical approach appears desirable.

In the following, we study the information dynamics with the help of the mutual information, $\mc{I}$. Since we have shown previously  \cite{touil2020quantum} that the OTOC sets a lower bound on the growth of $\mc{I}$, we can directly relate our findings with results from the literature. Specifically, in the present analysis we identify the additive contributions in the change of mutual information $\Delta \mc{I}$ corresponding to intrinsic scrambling and decoherence. This separation of terms permits us to derive an integral fluctuation theorem \cite{Deffner2019book} for scrambling under decoherence. Thus, as a main result, we obtain a statement of the second law of thermodynamics for quantum information scrambling. 

To gain further insight and build intuition into the dynamics of scrambling,  our conceptual analysis is complemented with the numerical analysis of five models. These models were specifically chosen to cover a wide range of scenarios and physical circumstances. The first two systems are quantum models that have holographic duals \cite{maldacena1999large}, where we get fast scrambling of quantum information. Specifically, we study the Sachdev-Ye-Kitaev (SYK) model, and two weakly coupled SYK-models through a traversable wormhole also known as the Maldacena-Qi (MQ) model.  Quantum gravity models are yet to be implemented experimentally \cite{chew2017approximating,chen2018quantum,marino2019cavity,lewis2019unifying,alavirad2019scrambling,
bentsen2019treelike,yin2020bound}. Thus, our numerical study may provide the groundwork for potential experiments and observations in the presence of decoherence. The remaining three models are variations of spin chain systems, where the goal is to further probe the role of chaos, ergodicity, and integrable dynamics in the scrambling of information in open spin chains.  The models are the disordered XXX-model in its ergodic phase, the disordered Mixed Field Ising (MFI) model, and the Lipkin-Meshkov-Glick (LMG) model.

\section{Stochastic thermodynamics of information scrambling}
\label{sec1}

In the first part of the analysis we develop the general framework and formulate statements of the second law of thermodynamics for scrambling. To this end, we consider a quantum system $\mc{S}$ with scrambling dynamics and $\mrm{dim}(\mathcal{S})=2^{N_\mathcal{S}}$. This system is  coupled to an environment $\mathcal{E}$, which induces noise that affects the scrambling of information in $\mc{S}$. The Hamiltonian of the universe then reads
\begin{equation}
\label{eq:ham}
H=H_\mc{S}\otimes\mbb{I}_\mc{E}+\mbb{I}_\mc{S}\otimes H_\mc{E}+h_\gamma\,,
\end{equation}
where $h_\gamma$ is  considered ``small''. In particular, we assume that the interaction energy is small enough that no energy is ``lost'' in the interaction, and that we have for the heat in $\mc{S}$ and $\mc{E}$, $\la Q_\mc{S}\ra\simeq-\la Q_\mc{E}\ra\equiv \la Q\ra $.

\subsection{Quantifying quantum information scrambling}

\paragraph*{Preliminaries.} Quantum information scrambling denotes the local loss of quantum information under unitary dynamics. In the conventional scenario, $\mc{S}$ is an isolated quantum system with $h_\gamma=0$, that is separated into two partitions $A$ and $B$. Now, imagine that a quantum state is prepared such that initially all information is concentrated in $A$, and the von Neumann entropy of $B$ is zero. Then, in scrambling dynamics the information leaks and eventually spreads throughout $B$, such that from any local measurements on $A$ the initial state can no longer be reconstructed.

\begin{figure}[h]
	\centering 
	\includegraphics[width=.48\textwidth]{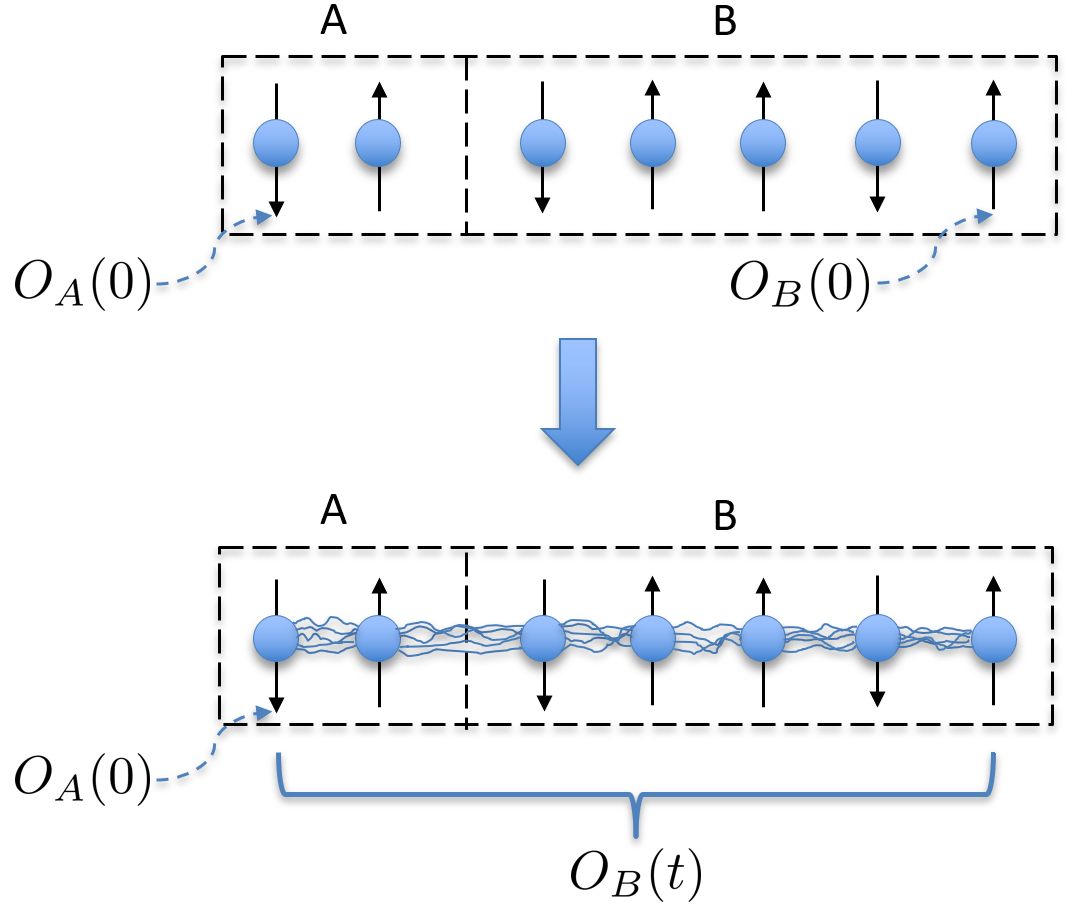}
	\caption{Illustration of the subsystem structure defining the support of the local operators $O_{A}$ and $O_{B}$, in a given spin chain. Under a scrambling unitary, the operator $O_{B}$ becomes highly non-local and spreads into the support of $A$, triggering the decay of the OTOC.}
	\label{chain}
\end{figure}

To measure this effect various quantifiers have been proposed in the literature \cite{swingle2018unscrambling,iyoda2018scrambling,Chenu2019workstatistics,touil2020quantum}. Arguably, the most common quantifier is the OTOC, which is a four-point correlation function that measures the operator growth in the Heisenberg picture,
\begin{equation}
\label{eq:otoc}
\mathcal{O}(t)=\la O_{A}^{\dagger} O_{B}^{\dagger}(t) O_{A} O_{B}(t)\ra\,,
\end{equation}
where the operators $O_{A}$ and $O_{B}$ initially act on different supports, as depicted in Fig.~\ref{chain}. 

Despite its ubiquitous use in the literature on information scrambling, the OTOC may not be the optimal quantity for thermodynamic analyses of open systems. For instance, the ambiguous choice of operators $O_{A}$ and $O_{B}$ makes the measure somewhat arbitrary. Therefore, one typically has to take averages in operator space \cite{touil2020quantum,yan2020information} which can become technically challenging. In addition, the OTOC alone cannot distinguish between the local effects of scrambling and decoherence \cite{landsman2019verified}. 

Thus, in the present analysis we build on our previous work \cite{touil2020quantum}. In fact, for a closed quantum system $\mathcal{S}$, under ``entangling unitary controls'',  we showed that the mutual information
\begin{equation}
 \mathcal{I} \equiv\mathcal{I}(A:B)=S_A+S_B-S_{AB}\,,
\end{equation}
with $S=-\tr{\rho\lo{\rho}}$, is  a good quantifier of information scrambling \cite{touil2020quantum}.  Specifically for unitary scrambling the change of mutual information $\Delta \mc{I}=\mc{I}(t)-\mc{I}(0)$ becomes \cite{touil2020quantum}
\begin{equation}
\label{eq:mut_info}
\Delta \mathcal{I}=\Delta S_{A}+\Delta S_{B}\,,
\end{equation}
which naturally lends itself to a thermodynamic analysis. Hence, it is a natural choice to employ the mutual information to also study scrambling in open systems.

\paragraph*{Entropy production in scrambling dynamics.}

We now fully turn our attention to open system dynamics. As $\mathcal{S}$ interacts with the environment $\mathcal{E}$ two physically distinct processes occur simultaneously: the scrambling of information within $\mathcal{S}$, and the destruction of coherences (and other quantum correlations) in $\mc{S}$ induced by $\mc{E}$. Thermodynamically, we can observe and quantify the interplay of these processes through different contributions to the entropy production.

To this end, we apply and generalize a framework that has been established in stochastic thermodynamics \cite{Esposito_2010}. For the sake of simplicity and without loss of generality, we assume that the global state of $\mc{S}$ and $\mc{E}$ is initially prepared, such that $\rho(0)=\rho_{\mathcal{S}}(0)\otimes \rho^\mrm{eq}_{\mathcal{E}}$, where $\rho^\mrm{eq}_{\mathcal{E}}=\e{-\beta H_\mc{E}}/Z_\mc{E}$ is the thermal Gibbs state of $\mc{E}$ at inverse temperature $\beta$. 

In general, the change of mutual information in $\mc{S}$  \eqref{eq:mut_info} becomes,
\begin{equation}
\label{eq:defI}
\Delta \mathcal{I}=\Delta S_{A}+\Delta S_{B}-\Delta S_\mc{S}\,,
\end{equation}
which accounts for the fact that the reduced dynamics of $\mc{S}$ is no longer unitary. Now exploiting that the global dynamics is unitary, and hence the von Neumann entropy of the universe, $\mc{S}\otimes\mc{E}$, is constant, and employing the usual tricks of adding and subtracting terms \cite{Esposito_2010} we obtain
\begin{equation}
\label{eq:delatI}
\Delta \mathcal{I}=\Delta S_{A}+\Delta S_{B}-D(\rho(t) || \rho_{\mathcal{S}}(t)\otimes \rho^\mrm{eq}_{\mathcal{E}} )-\Delta S_\mrm{ex}\,.
\end{equation} 
The first two terms in Eq.~\eqref{eq:delatI} describe the entropy production in each partition, which is analogous to scrambling in the isolated case \eqref{eq:mut_info}. The third term in Eq.~\eqref{eq:delatI} is the relative entropy measuring the buildup of correlations between system and environment,
\begin{equation}
\begin{split}
&D(\rho(t) || \rho_{\mathcal{S}}(t)\otimes \rho^\mrm{eq}_{\mathcal{E}} )=\tr{\rho(t)\lo{\rho(t)}}\\
&\qquad-\tr{\rho(t)\lo{\rho_{\mathcal{S}}(t)\otimes \rho^\mrm{eq}_{\mathcal{E}}}}\,,
\end{split}
\end{equation}
and $\Delta S_\mrm{ex}=\tr{(\rho_{\mathcal{E}}-\rho^\mrm{eq}_{\mathcal{E}}) \ln(\rho^\mrm{eq}_{\mathcal{E}})}\equiv\beta \langle Q \rangle$ is the change in excess entropy corresponding to the heat, $Q$, exchanged between $\mc{S}$ and $\mc{E}$ \cite{Esposito_2010}.

Now, further adding and subtracting the entropy of the environment, $\Delta S_\mc{E}$, we can write Eq.~\eqref{eq:delatI} as
\begin{equation}
\Delta \mathcal{I}=\Delta S_{A}+\Delta S_{B}- \mathcal{I(\mathcal{S}:\mathcal{E})}-D(\rho_{\mathcal{E}} || \rho^{eq}_{\mathcal{E}} )-\Delta S_\mrm{ex}\,.
\label{dar}
\end{equation}
The latter equality clearly exhibits the additive nature of the contributions to the entropy production of the distinct processes. This becomes even more transparent by using Eq.~\eqref{eq:defI} to re-write Eq.~\eqref{dar} as
\begin{equation}
\mathcal{I(\mathcal{S}:\mathcal{E})}+\Delta S_\mrm{ex}+D(\rho_{\mathcal{E}} || \rho^\mrm{eq}_{\mathcal{E}} )=\Delta S_\mc{S}\,.
\end{equation}
We recognize that any deviation from ideal, unitary scrambling will have three causes, namely (i) the buildup of correlations between $\mc{S}$ and $\mc{E}$, (ii) the exchange of heat between $\mc{S}$ and $\mc{E}$, and (iii) pushing $\mc{E}$ away from thermal equilibrium.

\subsection{Second law of thermodynamics for  scrambling}
\label{sec2}

To further the insight into the effect of environmental interference on scrambling, we need to clarify how $ \mathcal{I(\mathcal{S}:\mathcal{E})}$  is related to the other terms in Eq.~\eqref{dar}.  To this end, we will now derive an integral fluctuation theorem \cite{Deffner2019book}, i.e., a statement of the second law of thermodynamics for scrambling.

\paragraph*{General fluctuation theorem for open systems.}

Since we assume system and environment to be initially prepared in a product state, the change of mutual information between $\mc{S}$ and $\mc{E}$ becomes 
\begin{equation}
\begin{split}
\Delta \mathcal{I}(\mathcal{S:E})= \Delta S_{\mathcal{S}}+ \Delta S_{\mathcal{E}}\,.
\end{split}
\end{equation}
Therefore, a stochastic notion of entropy separated into contributions from $\mc{S}$ and $\mc{E}$ appears plausible.

For similar scenarios the two-time measurement approach \cite{campisi2011colloquium,deffner2011nonequilibrium,Kafri2012,Deffner2015PRL,Gardas2016,Talkner2016,Aberg2018,smith2018verification,Gardas2018} has proven convenient and powerful. In this paradigm, projective measurements on $\mc{S}$ and $\mc{E}$ are performed separately at $t=0$ and at some arbitrary later time $t=\tau$. Between $t=0$ and $t=\tau$ the universe is allowed to evolve under the Hamiltonian \eqref{eq:ham}. In its original inception \cite{campisi2011colloquium}, the two-time measurement scheme was developed to derive the quantum Jarzynski equality \cite{campisi2011colloquium,Talkner2016}. However, in more general settings the measurements can be chosen to suit the specific purposes \cite{deffner2011nonequilibrium,Kafri2012,Gardas2016}. 

For reasons that will become clear shortly, we choose the reduced Hamiltonian $H_\mc{S}$ to be the measured observable of $\mc{S}$, and we define
\begin{equation}
\label{eq:ent1}
\omega^{\mu \rightarrow\nu}_{\mathcal{S}}=\ln(p_\mc{S}^\mu)-\ln(p_\mc{S}^\nu),
\end{equation}
where $\mu$ is the outcome of the initial, and $\nu$ of the final energy measurement, and
\begin{equation}
p_\mc{S}^\mu=\bra{\mu} \rho_{\mathcal{S}}(0)\ket{\mu}\quad\text{and}\quad  p_\mc{S}^\nu=\bra{\nu}\rho_{\mathcal{S}}(t)\ket{\nu}\,.
\label{tomo1}
\end{equation}
Note that $\omega^{\mu \rightarrow\nu}_{\mathcal{S}}$ is nothing else but the stochastic entropy production of the Shannon entropy in energy representation \cite{Polkovnikov2011}. Generally, $\mc{S}$ is in a state far from thermal equilibrium at $t=0$ as well as at $t=\tau$. Thus, $\omega^{\mu \rightarrow\nu}_{\mathcal{S}}$ is not identical to a microscopic value of heat or work.

Since $\mc{E}$ is initially prepared in thermal equilibrium, an energy measurement is equivalent to state tomography. Given that we are after analyzing the buildup of correlations between $\mc{S}$ and $\mc{E}$ we choose the measurement at $t=\tau$ also to be state tomography on $\mc{E}$. Hence, we define
\begin{equation}
\label{eq:ent2}
\omega^{m\rightarrow n}_{\mathcal{E}}=\ln(p_\mc{E}^m)-\ln(p_\mc{E}^n)\,,
\end{equation}
where $p_\mc{E}^m$ and $p_\mc{E}^n$ are the initial and final eigenvalues of $\rho_\mc{E}$, respectively. Note that this definition is analogous, to the choice of measurements in Ref.~\cite{deffner2011nonequilibrium}. The difference is that in Ref.~\cite{deffner2011nonequilibrium} the energy of the environment is measured, whereas state tomography needs to be performed on the system.

We also would like to emphasize that at this point that the entropy production is defined entirely for conceptual reasons. From a practical point of view, suggesting state tomography on the environment is totally unreasonable. However, the purpose of the present part of the analysis is to study whether $\mc{E}$ can be used as a witness of scrambling in $\mc{S}$. Therefore, a more expensive measurement on $\mc{E}$ is inevitable. Further below, we will show how the requirement of full state tomography can be relaxed to situations in which only small fractions of $\mc{E}$ can be measured.

To continue we now consider the joint probability distribution of the stochastic entropy productions in $\mc{S}$ and $\mc{E}$.  As usual \cite{Deffner2019book} we write
\begin{equation}
\mc{P}(\omega_{\mc{S}}, \omega_{\mc{E}})=\la \delta (\omega_\mc{S}-\omega^{\mu \rightarrow\nu}_{\mathcal{S}})\,\delta(\omega_\mc{E}-\omega^{m\rightarrow n}_{\mathcal{E}})\ra\,,
\end{equation}
where the average is taken over the joint probabilities
\begin{equation}
\label{eq:joint}
p(\{\mu,m\}\rightarrow\{\nu,n\})=p(\nu,n|\mu,m)\,p^\mu_\mc{S}\,p^m_\mc{E}\,.
\end{equation}
Hence, we immediately obtain
\begin{equation}
\langle \omega_{\mathcal{S}} \rangle+\langle \omega_{\mathcal{E}} \rangle= \Delta \mathcal{I}(\mathcal{S:E})+ \Delta C(\mathcal{S}),
\label{hint}
\end{equation}
where $\mathcal{C}=S(\rho^\mrm{diag})-S(\rho)$ is the relative entropy of coherence in the energy eigenbasis \cite{baumgratz2014quantifying}, and $\rho^\mrm{diag}$ is the fully decohered diagonal matrix. Equation~\eqref{hint} serves as a justification of the choice of observables on $\mc{S}$ and $\mc{E}$. The average stochastic entropy production is given by the mutual information quantifying the correlations between system and environment, and the measure of the coherences present in $\mc{S}$. Moreover, Eq.~\eqref{hint} also demonstrates that in general the interaction with the environment not only destroys coherences, but also creates correlations (quantum and classical) between $\mc{S}$ and $\mc{E}$.

It is then easy to see that we also have
\begin{equation}
\label{eq:ft1}
\la \ex{-(\omega_{\mathcal{S}}+ \omega_{\mathcal{E}})}\ra= 1\,,
\end{equation} 
which follows from standard manipulations \cite{Deffner2019book} and the normalization of the joint probabilities \eqref{eq:joint}. Equation~\eqref{eq:ft1} constitutes an integral fluctuation theorem for general, open system dynamics. For any scenario in which a system $\mc{S}$ is prepared in a product state with an environment $\mc{E}$ the sum of the  stochastic  entropy productions in $\mc{S}$ and $\mc{E}$ fulfill an integral fluctuation theorem. Moreover, the measurements in $\mc{S}$ and $\mc{E}$ are chosen such that the average entropy production is given as a sum of correlations and coherences in $\mc{S}$.

 However, Eq.~\eqref{eq:ft1} cannot be regarded quite satisfactory for our present purposes. Demanding state tomography on $\mc{E}$ is only conceptually interesting, but practically unfeasible. In addition, Eq.~\eqref{eq:ft1} is a general result for open quantum systems without overly specific characteristics of information scrambling. Therefore, we continue the analysis by further refining the conceptual building blocks and derive further statements of the second law for scrambling dynamics.

\subsection{Environmental witness of scrambling}

We proceed to derive general bounds on the mutual information between $\mc{S}$ and $\mc{E}$. To this end, we now assume that $\tau\geq t^*$, where $t^*$ is the time at which $\mc{S}$ achieves maximal scrambling. In this case, Eq.~\eqref{dar} implies
\begin{equation}
\ex{\mathcal{I}(\mathcal{S}:\mathcal{E})} = 2^{N_\mathcal{S}} \ex{-\beta \langle Q \rangle-D(\rho_{\mathcal{E}} || \rho^{eq}_{\mathcal{E}} )}\,,
\label{general3}
\end{equation}
since  $\Delta S_{\mathcal{S}}=N_\mathcal{S}\ln(2)$ at maximal scrambling. Equation~\eqref{general3} shows that $\mc{I}(\mc{S}:\mc{E})$ can be computed from the heat exchanged between $\mc{S}$ and $\mc{E}$ and the relative entropy quantifying how far from equilibrium $\mc{E}$ is driven. Hence, the amount of information that $\mc{E}$ contains about the state of $\mc{S}$ is quantified by the heat exchanged between $\mc{S}$ and $\mc{E}$, and by how far $\mc{E}$ is pushed from equilibrium.

The motivation of the following arguments are similar to the conceptual underpinnings of \emph{Quantum Darwinism} \cite{zurek2009quantum,blume2006quantum}. Generally, observers do not need to access all of the environment's degrees of freedom in order to infer information about $\mathcal{S}$. For instance, as you read this paper, you intercept only a tiny fraction of the photons scattered off the physical paper, or emitted from a screen. This fraction of photons is enough to infer all the information contained in this paper. 

Thus, assuming that $\mc{E}$ is a true thermal reservoir, i.e., it remains in thermal equilibrium at all times, we rewrite Eq.~\eqref{dar} now as
\begin{equation}
\mathcal{I(S: P(E))}+\Delta S_\mrm{ex}\leq \Delta S_\mc{S}\,,
\label{dar1}
\end{equation}
where $\mathcal{P(\mathcal{E})}$ denotes a partition of $\mc{E}$. The latter inequality is a direct consequence of the strong subadditivity of the von Neumann entropy \cite{lieb1973proof}.

In particular, in the special case of purely decohering dynamics, $\Delta S_\mrm{ex}=0$, Eq.~\eqref{dar1} demonstrates the relation between the scrambling in $\mc{S}$ and the buildup of correlations between $\mc{S}$ and any partition of $\mc{E}$. Hence, observing only fractions of $\mc{E}$ may still give qualitative insight into the dynamics of $\mc{S}$, as the mutual information between the observed fraction and $\mc{S}$ sets a lower bound on the change of entropy in $\mc{S}$. Note that the bound becomes tight when $\mathcal{P(\mathcal{E}})=\mathcal{\mathcal{E}}$, or at the quantum to classical transition when measuring a partition of the environment is enough to infer all the information about the system \cite{zurek2009quantum,blume2006quantum}.

In summary, from Eq.~\eqref{dar} we derived an equality between the mutual information and a nonequilibrium thermodynamic quantity, which obeys an integral fluctuation theorem. Moreover, we derived bounds on the deviation from ideal information scrambling in determining  correlations between $\mc{S}$ and $\mc{E}$, as well as the amount of information we can learn by direct access to the environment, or partitions of the environment, with the latter representing the most physical scenario.

\section{Information scrambling and decoherence in the reduced dynamics}

\label{sec3}

In the previous section we elucidated the effect of environmental interaction on information scrambling. In the following, we complement the conceptual arguments with the numerical analysis of five models  with a wide range of physical characteristics. In particular, we now slightly change the point of view, and study the effects of decoherence on scrambling through the reduced dynamics of $\mc{S}$. This means, in particular, that we consider scenarios in which $\mc{E}$ is not experimentally accessible.

\subsection{Master equation for decohering dynamics}

We describe the interaction between $\mathcal{S}$ and $\mathcal{E}$, in the ultra weak coupling regime ($h_\gamma\ll 1$) by the master equation \cite{smith2018verification}
\begin{equation}
\frac{\pd \rho_\mc{S}}{\pd t}=-\frac{i}{\hbar}\,\com{H_\mc{S}}{\rho_\mc{S}}-\sum_{i\neq j} \gamma_{i j}\, \bra{i}\rho_\mc{S}\ket{ j}\ket{i}\bra{j},
\label{master1}
\end{equation}
where $\{|i\rangle\}_{i\in \llbracket 1,N \rrbracket}$ forms the decoherence basis, we drop the subscript ``$\mc{S}$'' when referring to the degrees of freedom of the system. Note that the first term in \myeqref{master1} is the unitary part governing the scrambling in $\mathcal{S}$, and the second term describes the interaction with $\mc{E}$. For the sake of simplicity, we further set $\gamma_{i j} \equiv \gamma$ (i.e. we suppress all the off-diagonal terms with the same rate), and we map our problem from the $N$-dimensional Hilbert space to the corresponding $N^2$-dimensional Fock-Liouville space. We relegate the mathematical details of the master equation in Fock-Liouville space to \aref{a}. For later reference, note that all the models of the following discussion live in a Hilbert space of dimension $\mrm{dim}(\mc{S})=2^6$, i.e. a 6-spin or 12-Majorana fermion system. 

Equation~\eqref{master1} and its derivation has been discussed in detail by Smith \etal \cite{smith2018verification}. In particular, it was shown that for decoherence in energy basis Eq.~\eqref{master1} follows from the quantum detailed balance master equations \cite{smith2018verification,alicki1976detailed}, by suppressing thermally induced transitions between energy eigenstates. Moreover, it is not hard to see that Eq.~\eqref{master1} describes unital dynamics \cite{smith2018verification}, and hence the quantum Jarzynski equality holds \cite{Kafri2012}.

For our present purposes, we consider two scenarios: (a) the dynamics induce scrambling, dissipation, and decoherence, (b) the dynamics induce only scrambling and decoherence. For the second case, i.e., for so-called \emph{pure decoherence} $\{|i\rangle\}_{i\in \llbracket 1,N \rrbracket}$ is composed of instantaneous eigenvectors of $H_\mc{S}(t)$. In this case, $\la H_\mc{S}(t)\ra=\text{const.}$, and hence no heat is exchanged between $\mc{S}$ and $\mc{E}$.

For scenario (a) we consider decoherence in the \emph{computational basis}, i.e., we take each vector $\ket{i} $ as the null vector except the $i$th entry with a value equal to 1. For instance, for a single qubit we have
\begin{equation}
\label{eq:comp}
\{|i\rangle\}_{i\in \llbracket 1,2 \rrbracket}=\left\{\begin{pmatrix} 1 \\ 0\end{pmatrix}\ ,\begin{pmatrix} 0 \\ 1\end{pmatrix}\right\}.
\end{equation}
In this case, the internal energy of $\mc{S}$ is no longer constant, and heat flows between $\mc{S}$ and $\mc{E}$.

Finally, in all following examples $\mc{S}$ is prepared either in the \emph{all-up} state
\begin{equation}
\label{eq:all-up}
\psi_\text{all-up}\propto\ket{00\dots 00},
\end{equation}
or in the \emph{N\'eel} state,
\begin{equation}
\label{eq:Neel}
\psi_\text{N\'eel}\propto\ket{0101\dots 10}\,.
\end{equation}
This choice is made, since it has been shown that for these states the following models exhibit information scrambling in the unitary case.

\subsection{High energy physics models}
\label{sec3b}

We start with two representative models that were formulated in the context of $AdS_2/CFT_1$. More mundanely, the chosen models can be understood as 1-D quantum models, through a straightforward Jordan-Wigner transformation of the fermionic field operators \cite{batista2001generalized,cotler2019spectral,touil2020quantum}, or as 2-D black hole models in anti-de Sitter space \cite{bengtsson1998anti}.

\begin{figure}
	\centering 
	\includegraphics[width=.44\textwidth]{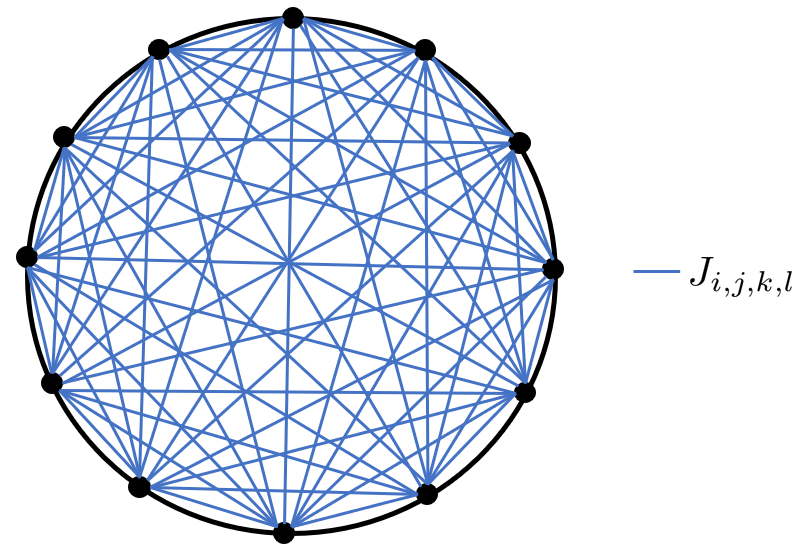}
	\caption{Sktech of the SYK-model with 12 Majorana fermions. The network represents the all-to-all interactions between the fermionic sites.}
	\label{SYK}
\end{figure}

\paragraph*{Sachdev-Ye-Kitaev model.} As a first example, we study the Sachdev-Ye-Kitaev (SYK) model, which is a quantum gravity vector model of $N$ strongly interacting Majorana fermions \cite{Rosenhaus_2019,PhysRevD.94.106002,PhysRevLett.70.3339}, see a depiction in Fig.~\ref{SYK}.  From a high energy perspective, and in the limit of large $N$, it models the scrambling properties of a 1+1-D black hole in anti-de Sitter space, and hence it has found applications in solutions to the information paradox \cite{paradox1,mathur2009information,preskill1992black}. From a quantum information theoretic perspective the SYK-model is nothing but a collection of qubits that chaotically scramble information. For instance, the quantum chaotic behavior of the SYK-model, was investigated in the context of quantum batteries \cite{rossini2019quantum,rosa2019ultra}, showing the potentially practical nature of such models.  

The Hamiltonian of the SYK-model reads
\begin{equation}
H_\mrm{SYK}=- \sum_{1 \leq i_{1}<i_{2}<i_{3}<i_{4} \leq N} J_{i_{1} i_{2} i_{3} i_{4}} \psi_{i_{1}} \psi_{i_{2}} \psi_{i_{3}} \psi_{i_{4}},
\label{ham_syk}
\end{equation}
where we take four interacting Majorana fermions at each instant. Here, $J_{i_{1} i_{2} i_{3} i_{4}}$ are real independent random variables with values drawn from a Gaussian distribution with mean $\left\langle J_{i_{1} i_{2} i_{3} i_{4}}\right\rangle=0$ and variance $\left\langle J_{i_{1} i_{2} i_{3} i_{4}}^{2}\right\rangle=J^{2}(3) !/N^{3}$. Moreover, the $\psi_i$ are the field operators of the Majorana fermions.

\begin{figure*}
\begin{minipage}[l]{.48\textwidth}
Decoherence in computational basis\\
\vspace{.5em}
\includegraphics[width=\textwidth]{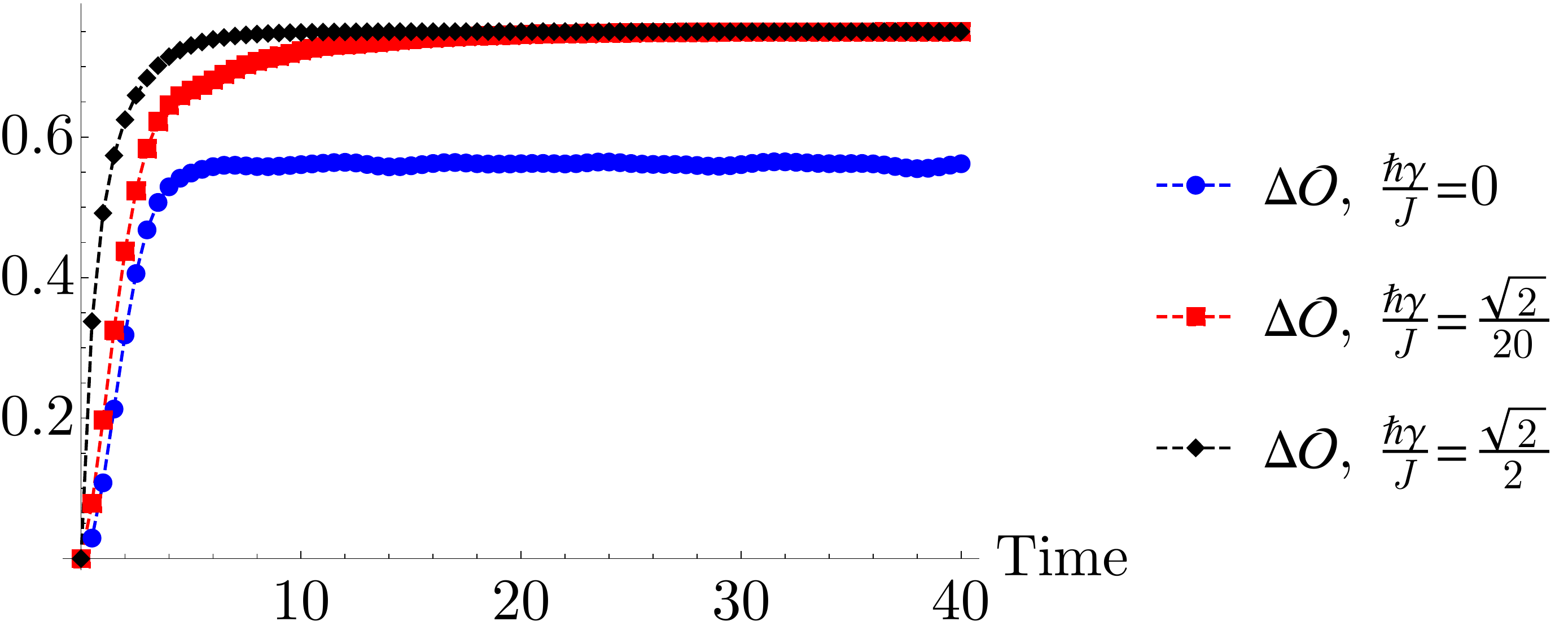}
\end{minipage}
\hfill
\begin{minipage}[l]{.48\textwidth}
Decoherence in energy basis\\
\vspace{.5em}
\includegraphics[width=\textwidth]{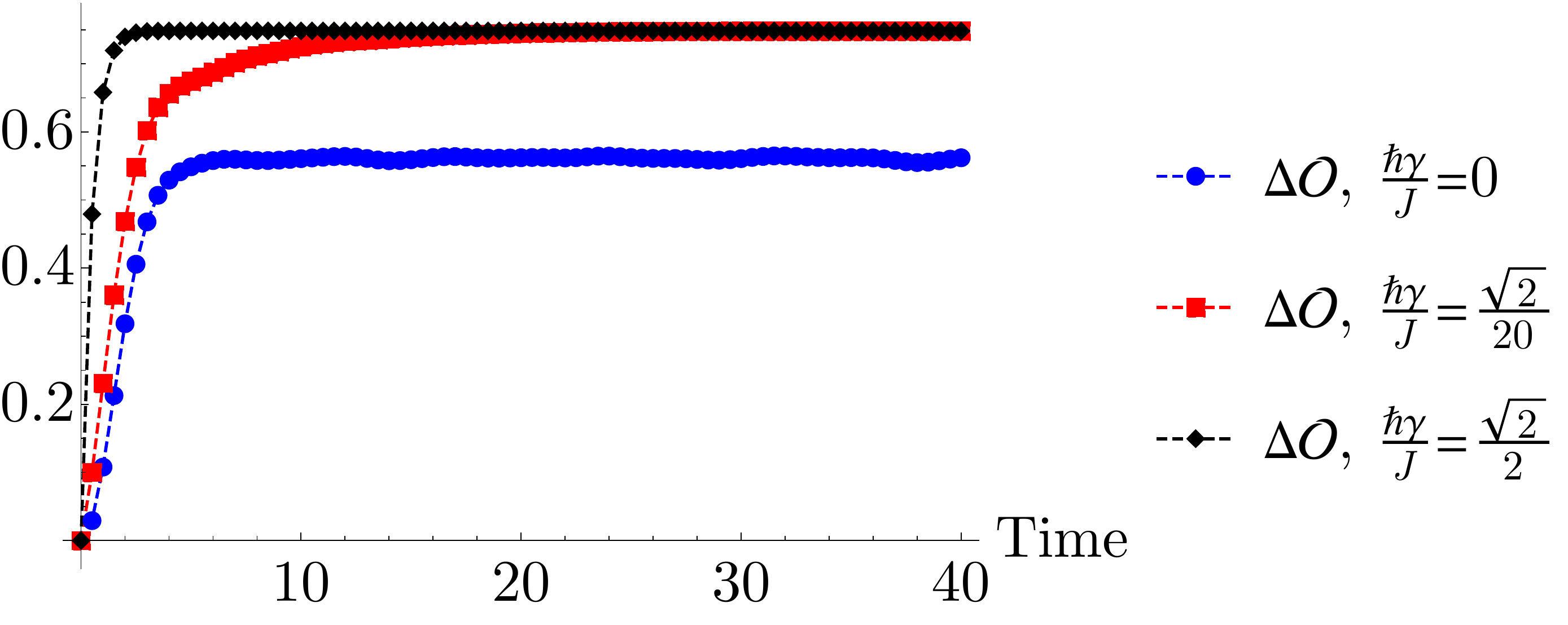}
\end{minipage}
\caption{Growth of the average OTOC \eqref{eq:otoc} as a function of time for the SYK-model with $N=12$ and an initial ``all-up'' state \eqref{eq:all-up}. Results were obtained as averages over $10^2$ realization.}
	\label{otoc}
\end{figure*}
\begin{figure*}
\begin{minipage}[l]{.48\textwidth}
Decoherence in computational basis\\
\vspace{.5em}
\includegraphics[width=\textwidth]{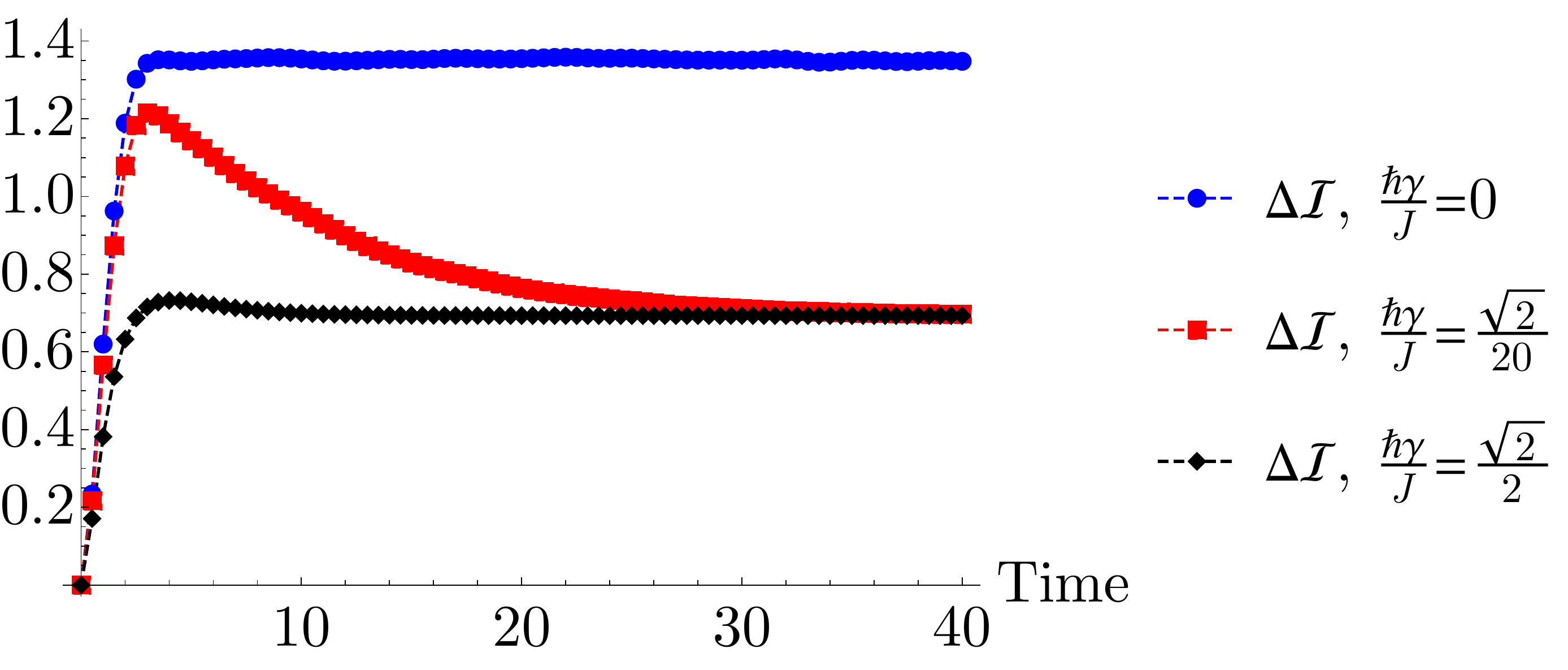}\\
\includegraphics[width=\textwidth]{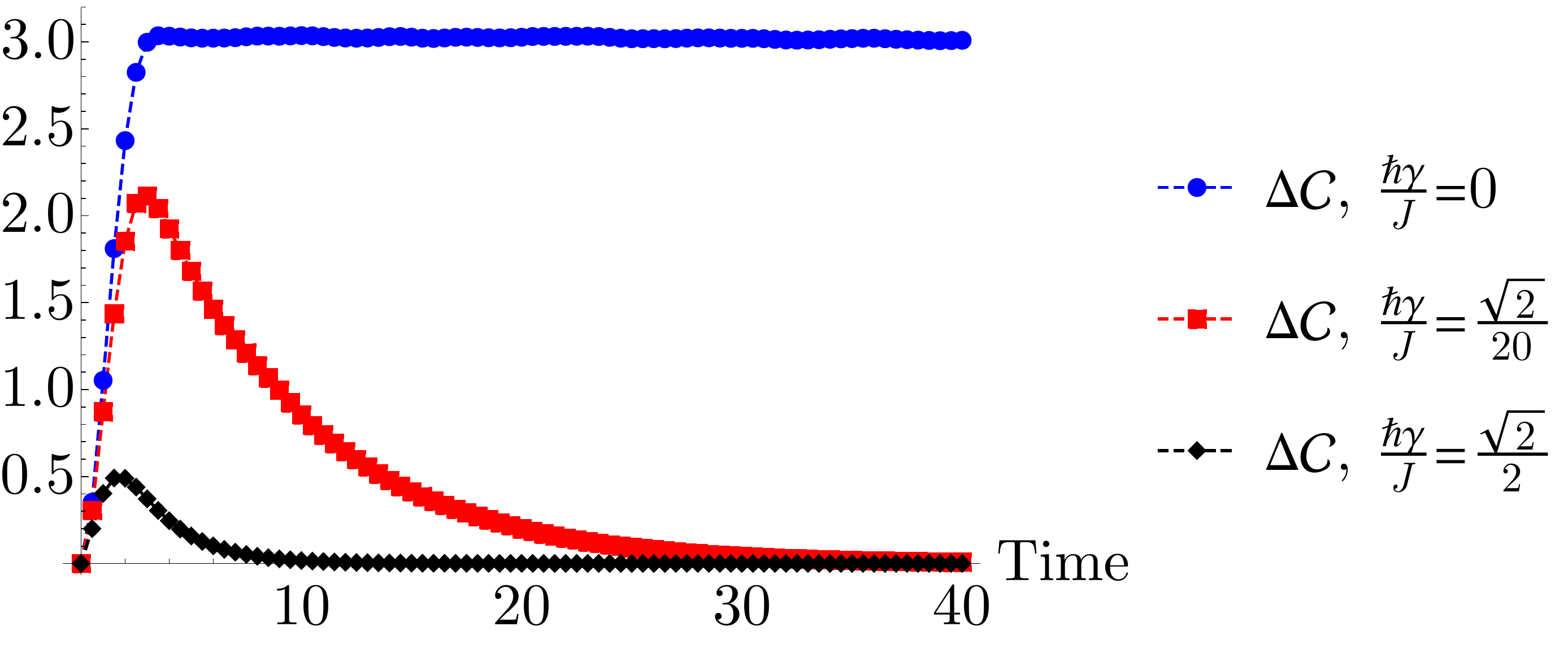}
\end{minipage}
\hfill
\begin{minipage}[l]{.48\textwidth}
Decoherence in energy basis\\
\vspace{.5em}
\includegraphics[width=\textwidth]{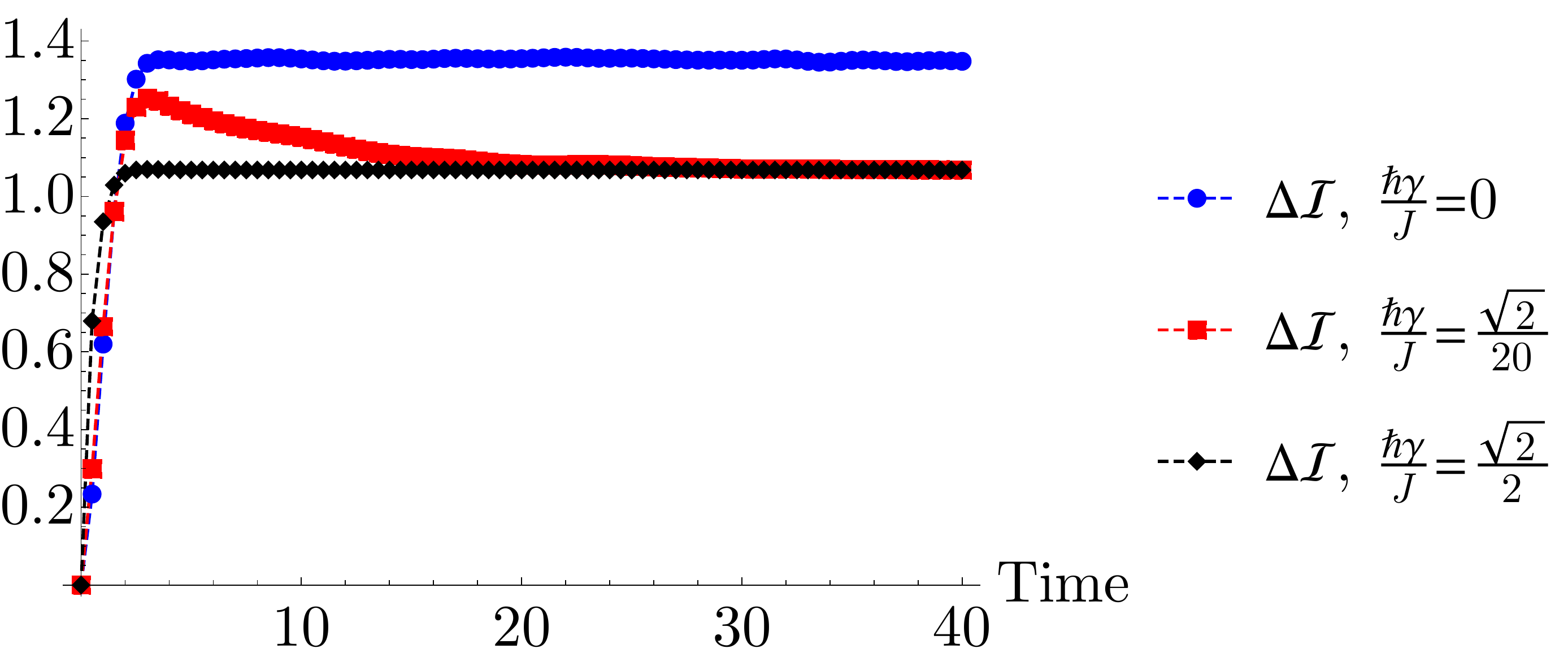}\\
\includegraphics[width=\textwidth]{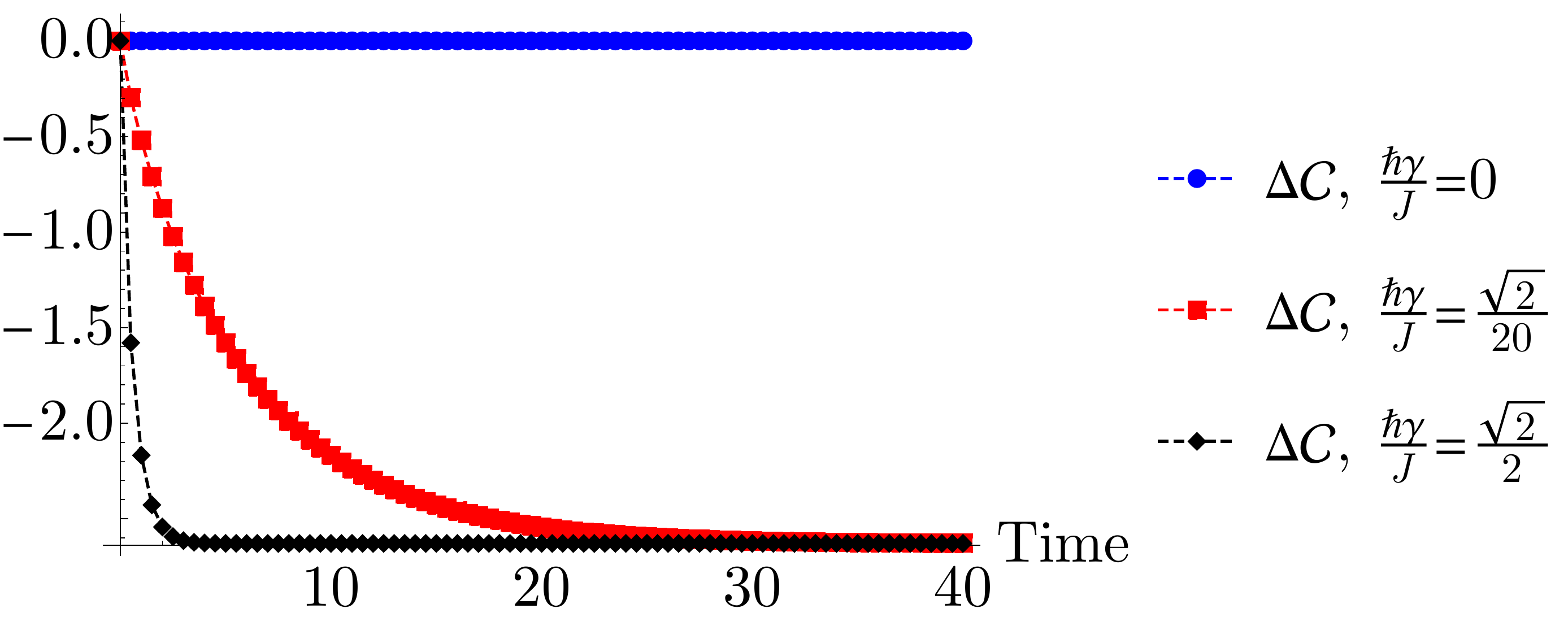}
\end{minipage}
\caption{Evolution of the change of mutual information ($\Delta \mathcal{I}$) and relative entropy of coherence ($\Delta \mathcal{C}$) for the SYK-model \eqref{ham_syk} with $N=12$, for an initial ``all-up'' state \eqref{eq:all-up}.  Results were obtained as averages over $3 \times 10^2$ realizations.}
	\label{fig1a}
\end{figure*}

We solved the dynamics of the SYK-model \eqref{ham_syk} as described by Eq.~\eqref{master1} for both, decoherence in the energy as well as decoherence in the computational basis. As initial state we chose the all-up state \eqref{eq:all-up}, for which the SYK-model exhibits fast scrambling for $\gamma=0$. Here and in the following, the partitions $A$ and $B$ are chosen such that $A$ is a single qubit and its complement in $\mathcal{S}$ represents $B=A^{C}$. Maximal scrambling is indicated by $\mathcal{I}=2\ln(2)$, and hence any environmental effect leads to $\mathcal{I} < 2 \ln(2)$ in the long time limit. It turns out that averaging over 10 random Hamiltonians gives sufficient convergence. For completeness, Fig.~\ref{otoc} depicts the resulting Haar averaged OTOC: $\Delta\mathcal{O}\equiv \int_{\text {Haar }} \left(\mathcal{O}(0)-\mathcal{O}(t)\right) \mathrm{d} O_A \mathrm{d} O_B$ for zero ($\hbar\gamma/J=0$), weak ($\hbar\gamma/J\ll1$), and medium ($\hbar\gamma/J\simeq 1$) coupling between $\mc{S}$ and $\mc{E}$.

We observe that the OTOC is not a good quantifier of scrambling for open systems, as there is no direct way to distinguish the closed system dynamics (where we indeed have scrambling) and the open system case where both scrambling and decoherence take effect. The latter remark applies for both cases of decoherence: in the energy as well as the computational basis,  we always get monotonically increasing functions with time.  Therefore, we focus on the change of the mutual information,  $\Delta \mathcal{I}\equiv \mathcal{I}(t)-\mathcal{I}(0)$, and the relative entropy of coherence, $\Delta \mathcal{C}\equiv \mathcal{C}(t)-\mathcal{C}(0)$. The results are summarized in Fig.~\ref{fig1a}.

For decoherence in the computational basis \eqref{eq:comp}, we observe that for weak coupling to $\mc{E}$ the mutual information reaches a maximum early in the evolution, and hence information is indeed  initially scrambled throughout $\mc{S}$. At later times, decoherence takes over and the mutual information reaches a stationary value that is independent of the coupling strength $\gamma$. Similarly, we see that initially coherences are built up, which are then inevitably destroyed in the open system dynamics.

The situation is similar, yet also markedly different for decoherence in the energy basis. The all-up state \eqref{eq:all-up} is actually maximally coherent (in energy representation), and the change of the relative entropy of coherence is negative. The behavior of the mutual information is similar to the one observed for decoherence in the computational basis. For weak coupling, information is initially scrambled, before the mutual information reaches a stationary value independent of the coupling strength $\gamma$. However, this stationary value is larger for decoherence in the energy basis than for decoherence in the computational basis. This can be understood by considering that for pure decoherence $\Delta S_\mrm{ex}=0$, and hence the deviation from ideal scrambling is given by only two, instead of three contributions, compare Eq.~\eqref{dar}.

\begin{figure}
	\centering 
	\includegraphics[width=.46\textwidth]{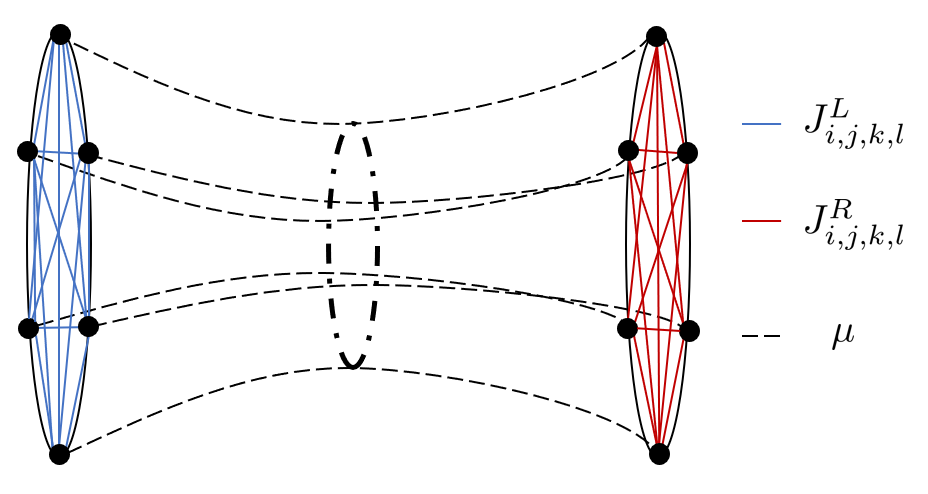}
	\caption{Sketch of the MQ-model with 12 Majorana fermions, and $J^{L}_{i,j,k,l}=J^{R}_{i,j,k,l}=J_{i,j,k,l}$.}
	\label{illustration}
\end{figure}

\paragraph*{Wormhole in Anti-de Sitter space.} As a second example, we study a more complex scenario. The Maldacena-Qi (MQ) model \cite{maldacena2018eternal,plugge2020revival} consists  of two weakly coupled SYK-models, and it can be interpreted as an eternal traversable wormhole in $AdS_2$ with two black holes at its sides. Its Hamiltonian reads
\begin{equation}
H_\mrm{MQ}=H^{L}_{\mathrm{SYK}}+H^{R}_{\mathrm{SYK}}+i \mu \sum_{j} \psi^{L}_{j} \psi^{R}_{j},
\label{MQe}
\end{equation}
where $H^{L}_{\mathrm{SYK}}$ and $H^{R}_{\mathrm{SYK}}$ describe the left and right black holes given by Eq.~\eqref{ham_syk}. Further, $\psi^{L}_{j}$ and $\psi^{R}_{j}$ are the left and right fermionic field operators, respectively. A sketch of the model can be found in Fig.~\ref{illustration}.

The MQ-model describes a wormhole due to the revival dynamics happening between the two black holes; a perturbation in one of the SYK black holes travels to the other side (the image black hole). For a detailed analysis on the revival dynamics or other intriguing properties of this model we refer to the literature~\cite{plugge2020revival,garcia2019quantum,maldacena2019syk,chen2019entanglement,alet2020entanglement}. 

For the present purposes, we are interested in the information scrambling dynamics of the model as a whole, i.e., we track the growth of entanglement between a single fermionic site and the rest of the sites in both black holes. To this end, we consider the weak coupling regime, $\mu \ll J$, and as before the initial state of the composite system is ``all-up'' \eqref{eq:all-up}.

\begin{figure*}
\begin{minipage}[l]{.48\textwidth}
Decoherence in computational basis\\
\vspace{.5em}
\includegraphics[width=\textwidth]{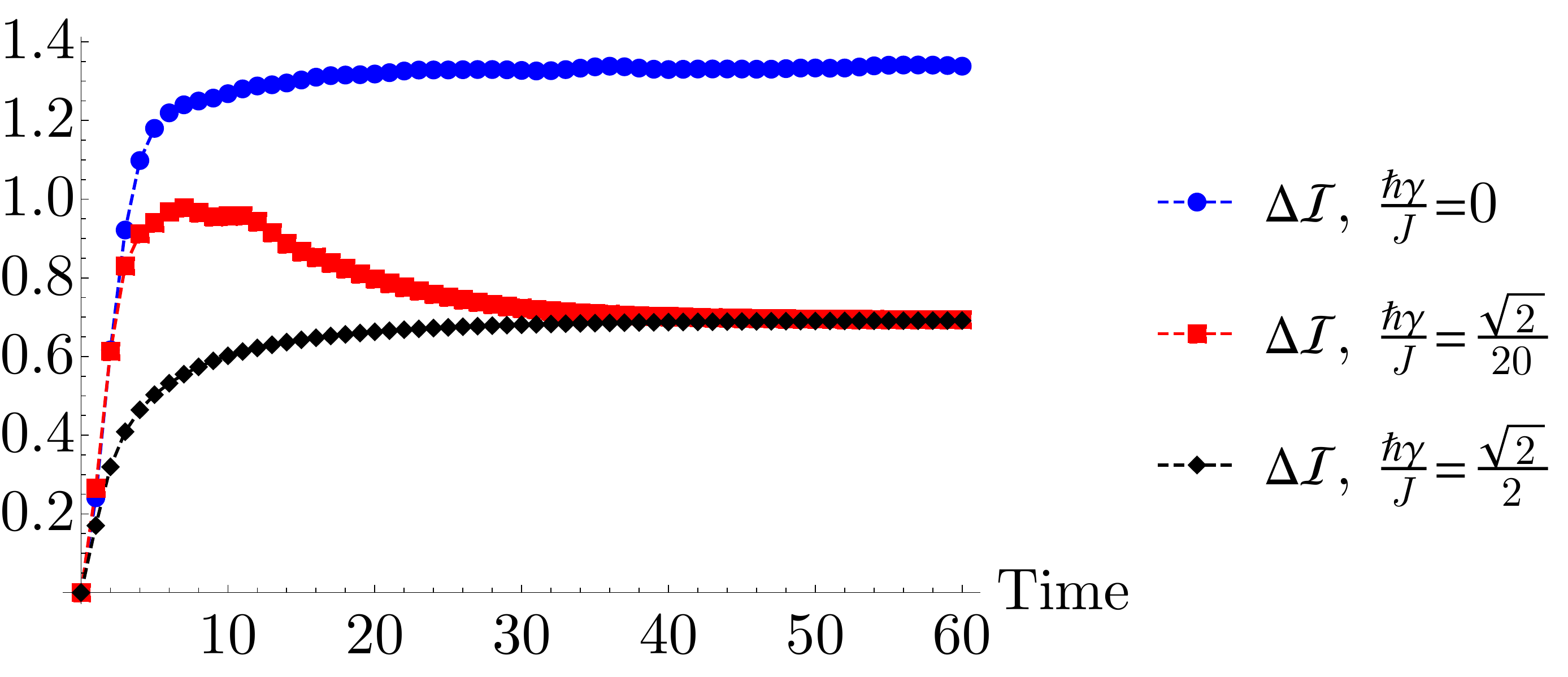}\\
\includegraphics[width=\textwidth]{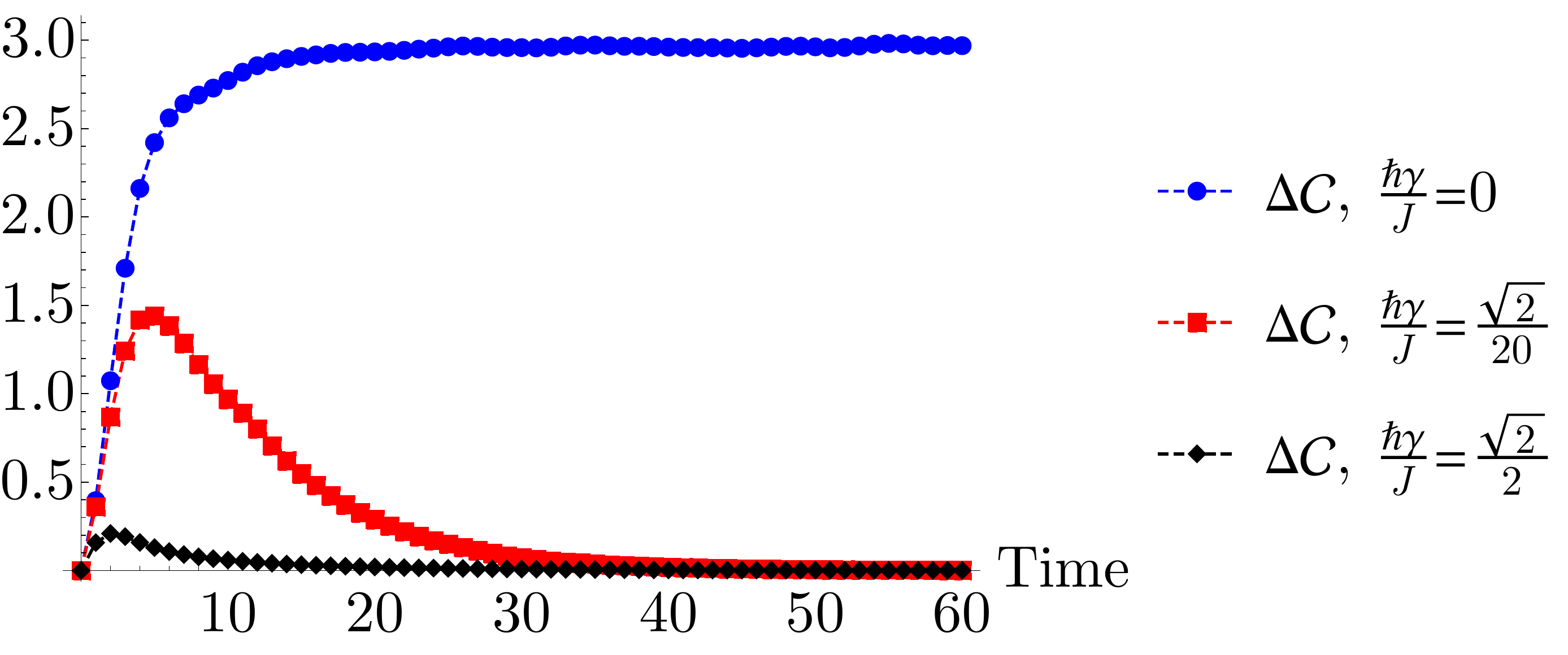}
\end{minipage}
\hfill
\begin{minipage}[l]{.48\textwidth}
Decoherence in energy basis\\
\vspace{.5em}
\includegraphics[width=\textwidth]{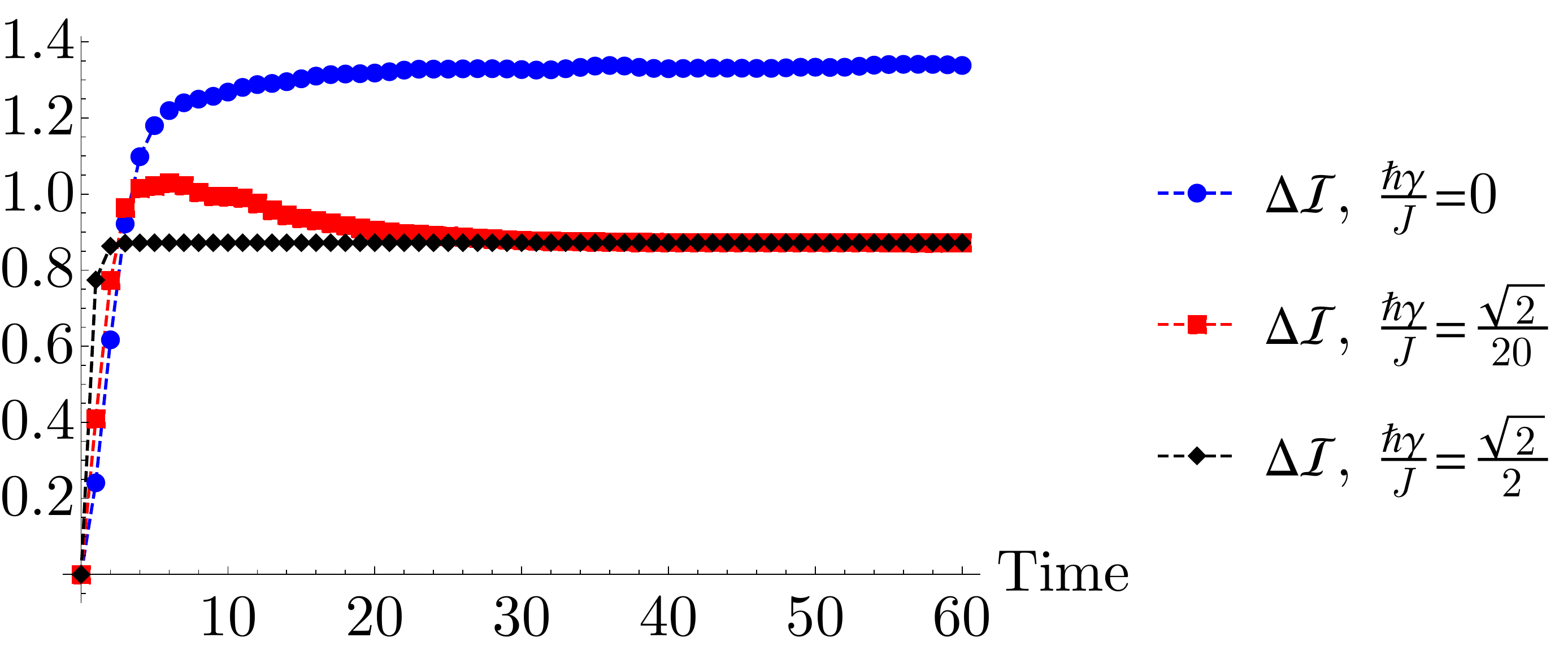}\\
\includegraphics[width=\textwidth]{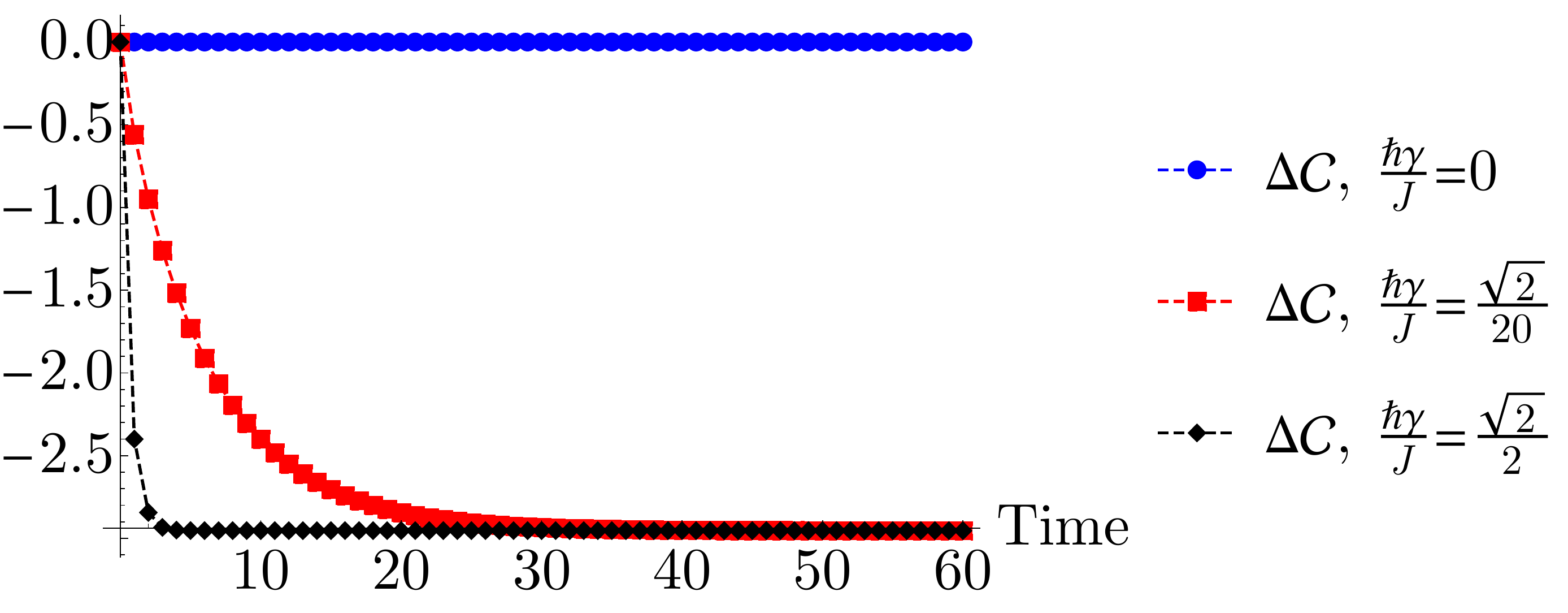}
\end{minipage}
\caption{Evolution of the change in mutual information ($\Delta \mathcal{I}$) and relative entropy of coherence ($\Delta \mathcal{C}$) for the MQ-model \eqref{MQe} with $\mu=0.1$. Each black hole is composed of 6 Majorana fermionsm which gives a total of 12 fermions living in a Hilbert space of $2^6$ dimensions.  Results were obtained as averages over $3 \times 10^2$ realizations.}
	\label{MQ}
\end{figure*}

Comparing the numerical findings for a single SYK-model in Fig.~\ref{fig1a} and for the MQ-model in Fig.~\ref{MQ} we observe strikingly similar behavior. Despite the significantly higher complexity of the MQ-model hardly any new insight into information scrambling in the presence of decoherence is obtained, beyond what we discussed above for the SYK-model. 

On a more speculative note, it seems glaringly obvious that if information scrambling is to play a fundamental role in the resolution of the information paradox then internal decoherence due to the gravitational fields cannot be neglected. However, such cosmological questions are somewhat beyond the scope of the present analysis. Therefore, we continue the analysis with the more mundane study of spin chain models.

\subsection{Spin chain models}
\label{sec4}

\begin{figure}
	\centering 
	\includegraphics[width=.46\textwidth]{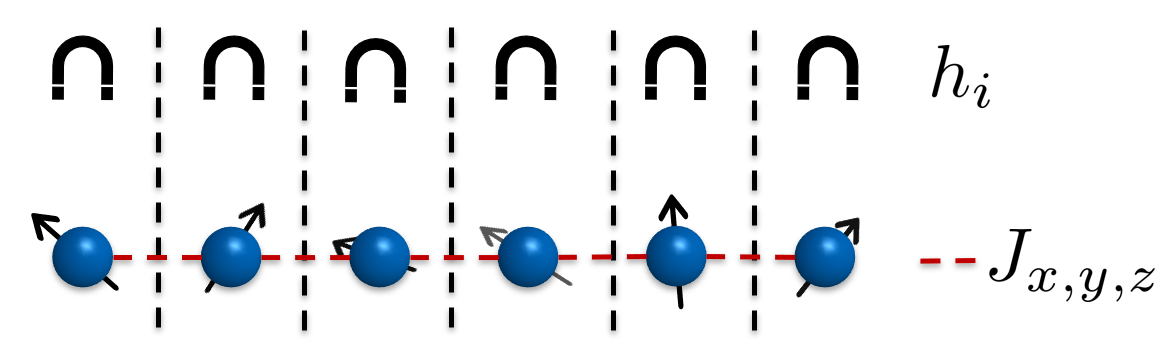}
	\caption{Sketch of the XXX-model with nearest neighbor interactions, and random local fields $h_i$.}
	\label{xxx}
\end{figure}

\paragraph*{Disordered XXX-model.} As the first case study of a  spin chain model we numerically analyze the disordered XXX-model. It has been shown in the literature \cite{iyoda2018scrambling} that this model is ergodic and exhibits information scrambling when initialized in the N\'eel state \eqref{eq:Neel}. The Hamiltonian reads
\begin{equation}
H_\mrm{XXX}=\sum_{\langle i, j\rangle} J \boldsymbol{\sigma}_{i} \cdot \boldsymbol{\sigma}_{j}+\sum_{i=1}^{N} h_{i} \sigma_{i}^{z}\,,
\label{x}
\end{equation}
where $\boldsymbol{\sigma}$ is the Pauli vector. The model is denoted  ``XXX'', since the three interaction coefficients are designed to be equal $J_x=J_y=J_z\equiv J$. The XXX-model is comprised of $N$ spin-1/2 particles with nearest neighbor interactions and random local transverse magnetic fields $h_i$. The model is depicted in Fig.~\ref{xxx}.

\begin{figure*}
\begin{minipage}[l]{.48\textwidth}
Decoherence in computational basis\\
\vspace{.5em}
\includegraphics[width=\textwidth]{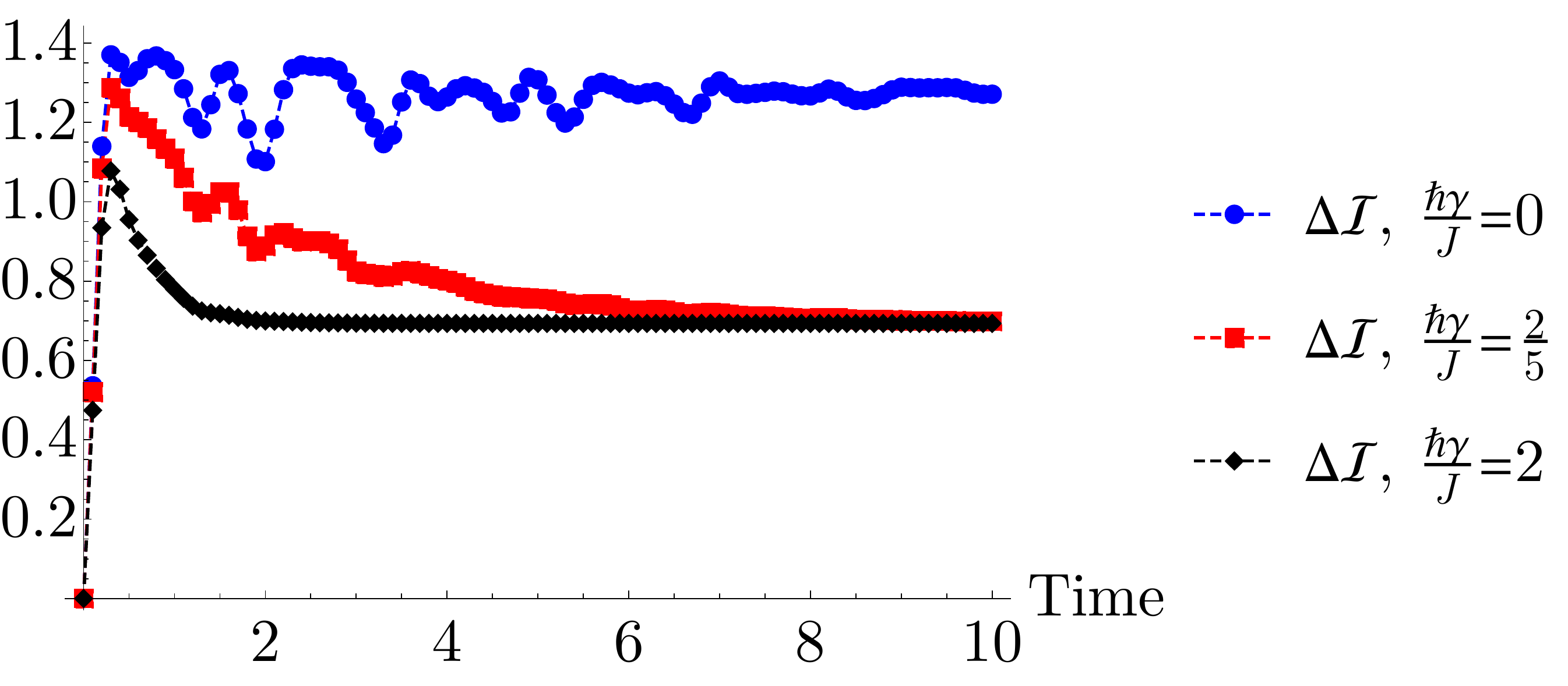}\\
\includegraphics[width=\textwidth]{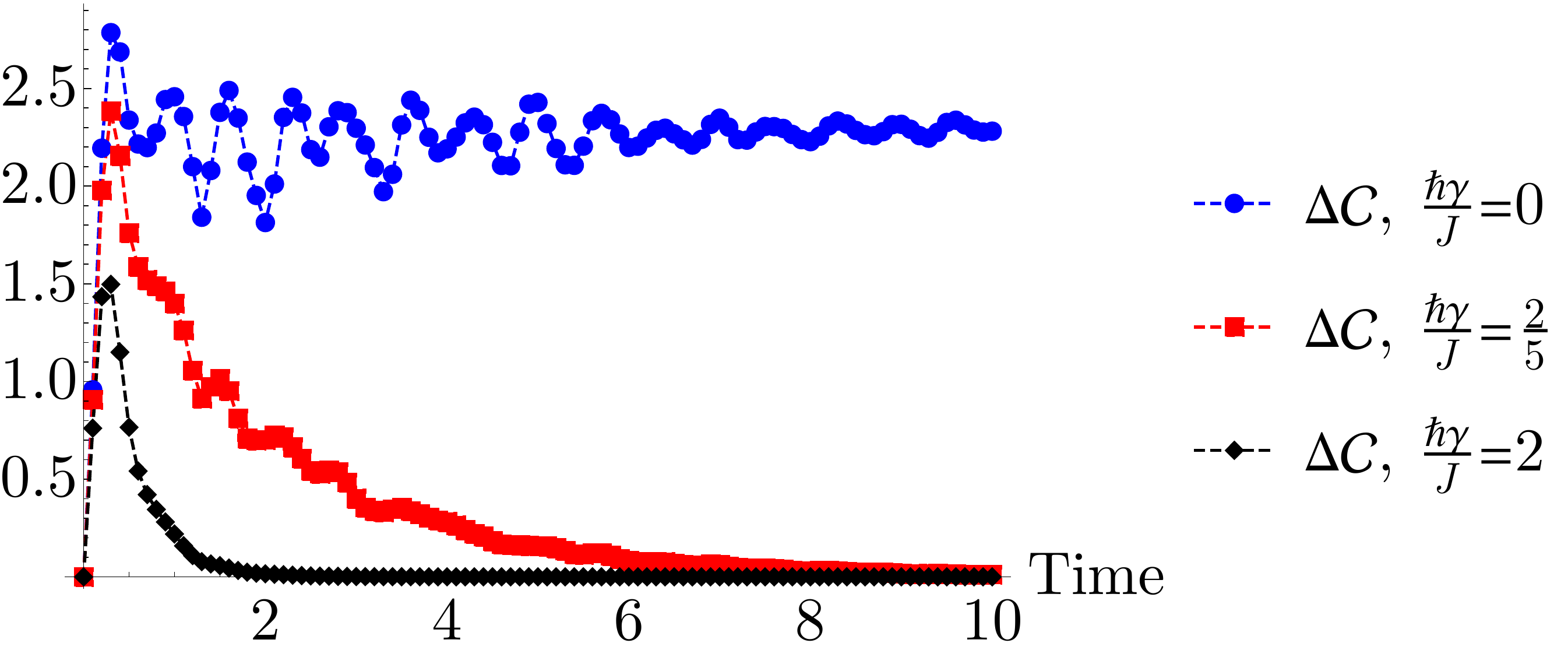}
\end{minipage}
\hfill
\begin{minipage}[l]{.48\textwidth}
Decoherence in energy basis\\
\vspace{.5em}
\includegraphics[width=\textwidth]{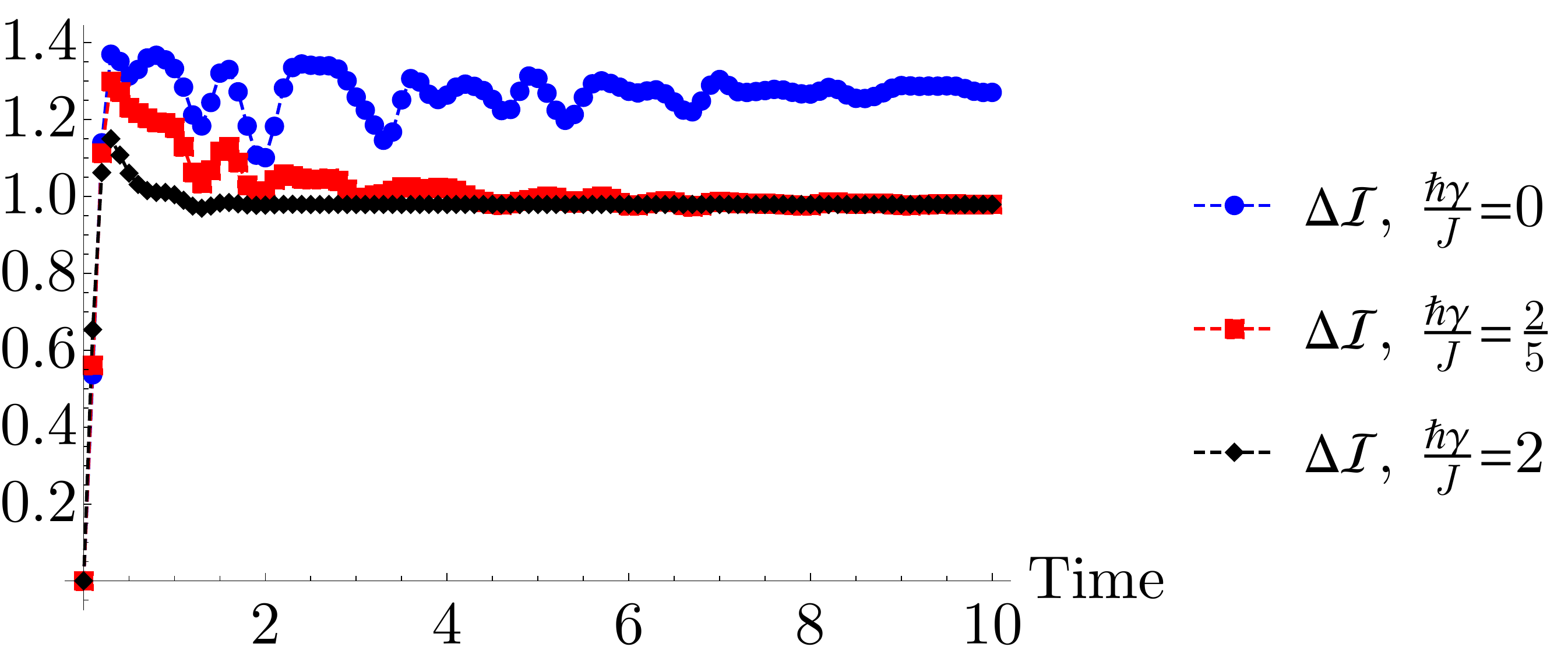}\\
\includegraphics[width=\textwidth]{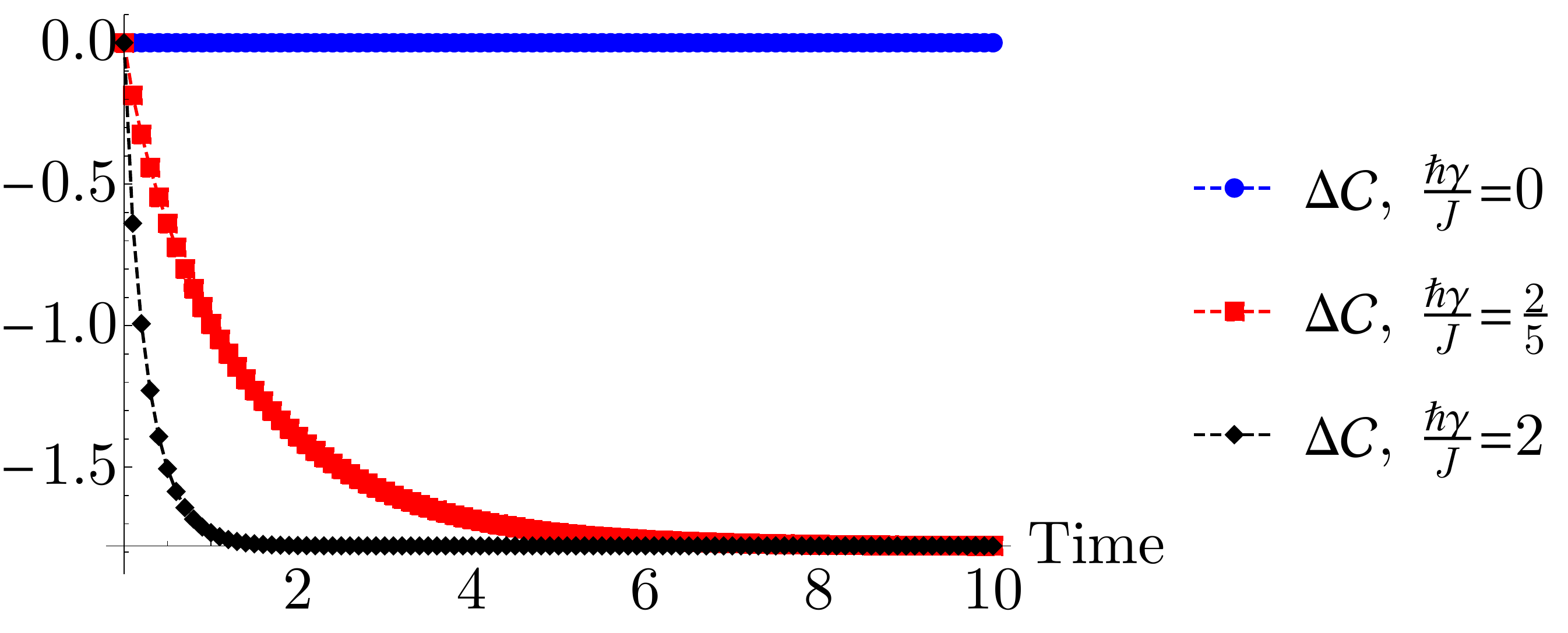}
\end{minipage}
\caption{Evolution of the change in mutual information ($\Delta \mathcal{I}$) and relative entropy of coherence ($\Delta \mathcal{C}$) for the disordered XXX-model \eqref{x} in its ergodic phase with $N=6$ and for an initial N\'eel state \eqref{eq:Neel}.  Results were obtained as averages over $4 \times 10^2$ realizations.}
	\label{xx}
\end{figure*}

For our present purposes, $h_{i}$ are drawn uniformly from $[-h,h]$, and we set $h=J=1$. Hence, the model is studied in its ergodic phase and we avoid the many body localized (MBL) phase that exhibits no scrambling of information \cite{iyoda2018scrambling}. Figure~\ref{xx} summarizes our findings.

We observe that the plots are significantly noisier than for the SYK- and the MQ-models. This behavior is similar to what has been reported in the literature \cite{iyoda2018scrambling}. Otherwise, the XXX-model in its ergodic phase exhibits the same qualitative features that we found for the SYK- and the MQ-model.

\paragraph*{Mixed field Ising model.}

\begin{figure}
	\centering 
	\includegraphics[width=.46\textwidth]{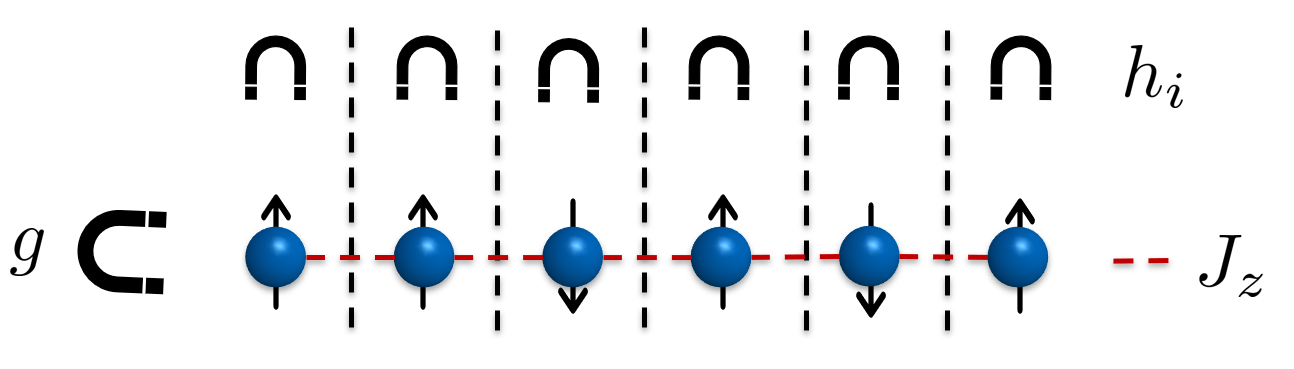}
	\caption{Sketch of the MFI-model with nearest neighbor interactions (in $z$-direction), random local fields $h_i$, and a global field in $x$-direction.}
	\label{mfi}
\end{figure}

\begin{figure*}
\begin{minipage}[l]{.48\textwidth}
Decoherence in computational basis\\
\vspace{.5em}
\includegraphics[width=\textwidth]{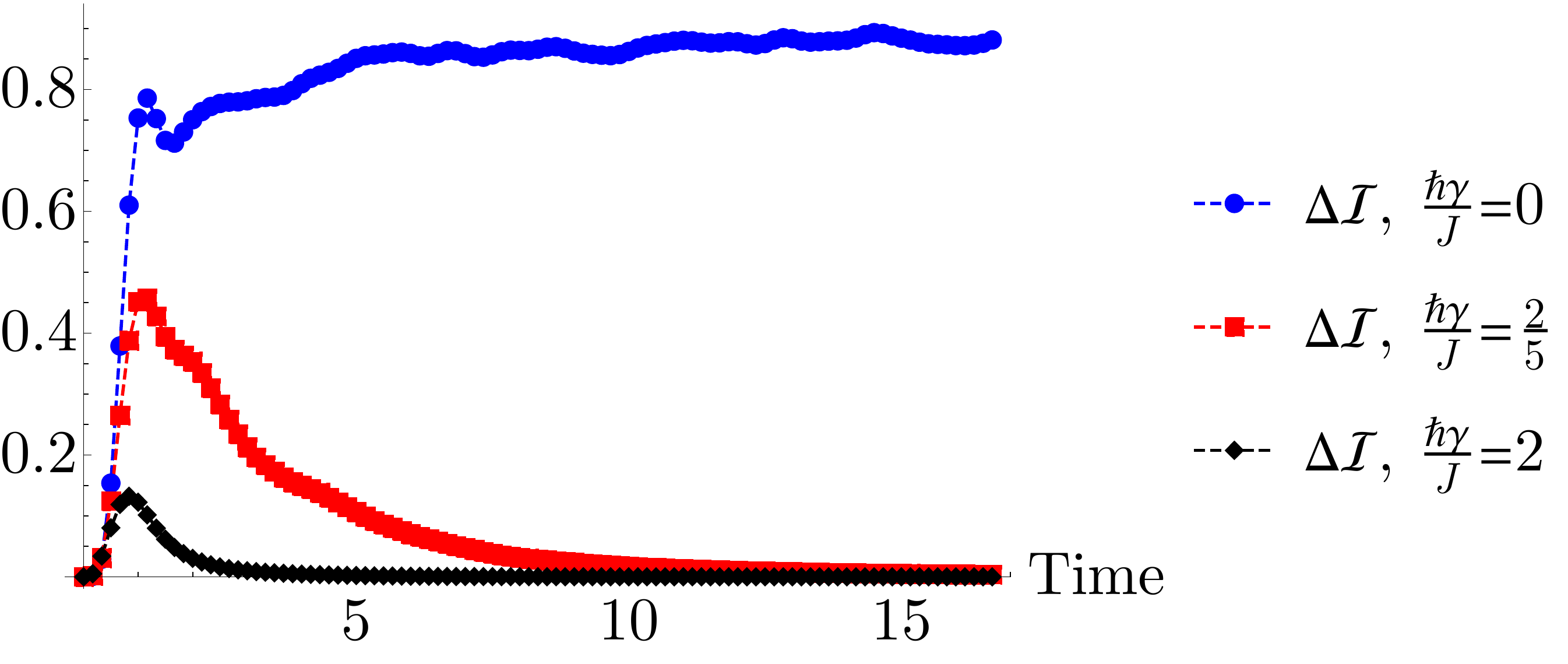}\\
\includegraphics[width=\textwidth]{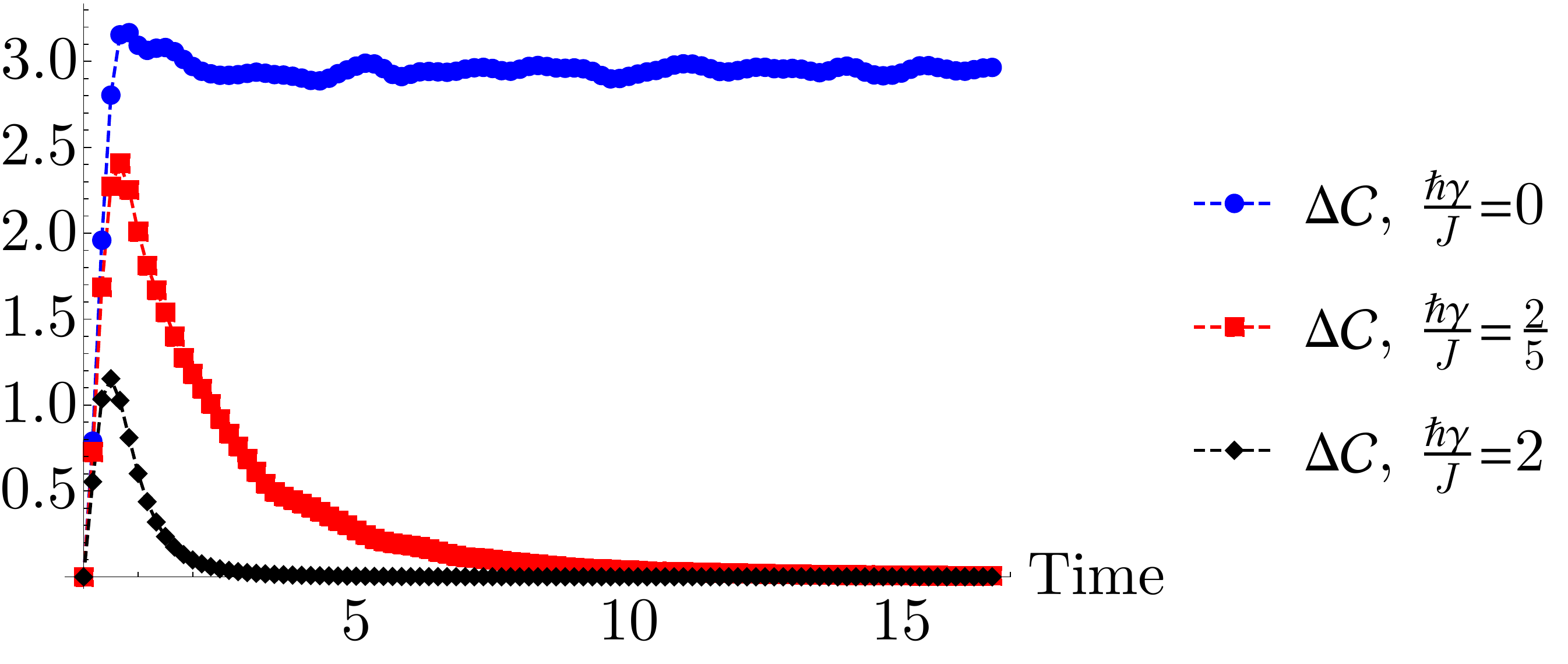}
\end{minipage}
\hfill
\begin{minipage}[l]{.48\textwidth}
Decoherence in energy basis\\
\vspace{.5em}
\includegraphics[width=\textwidth]{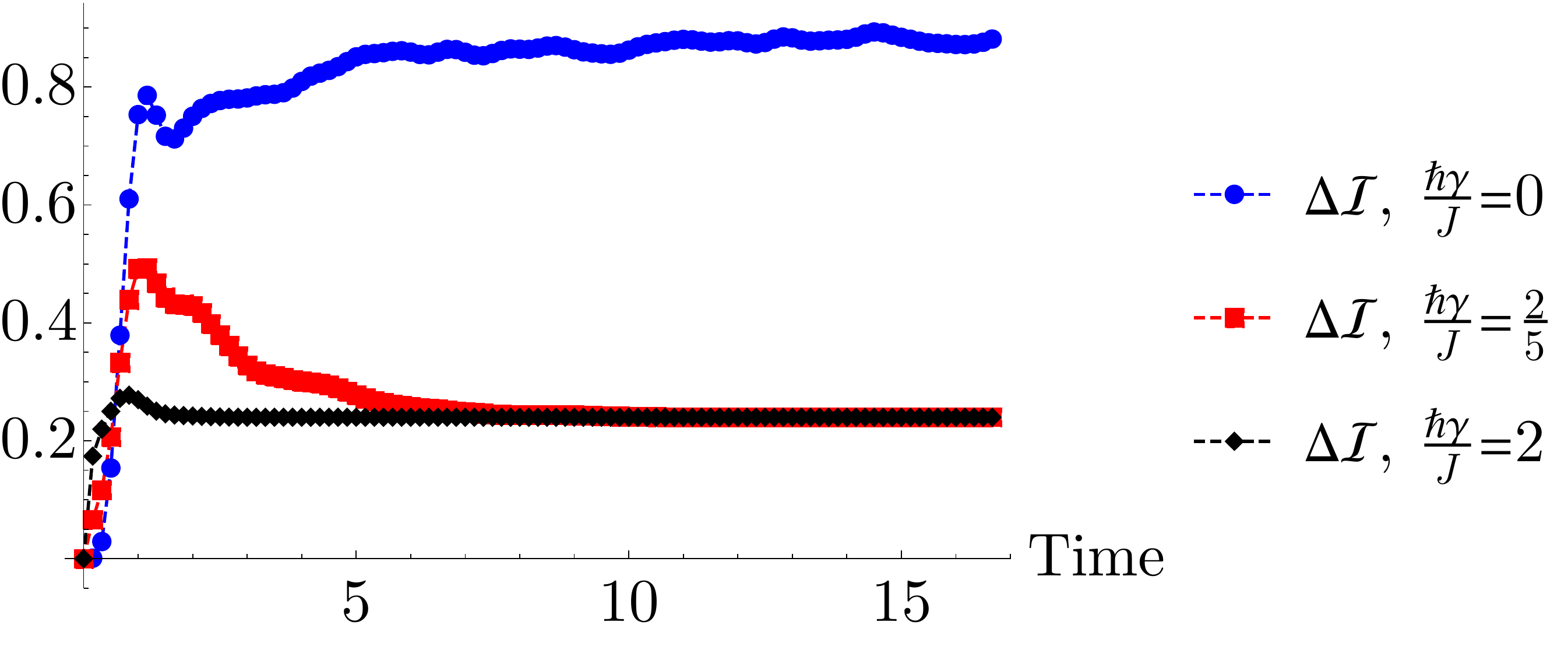}\\
\includegraphics[width=\textwidth]{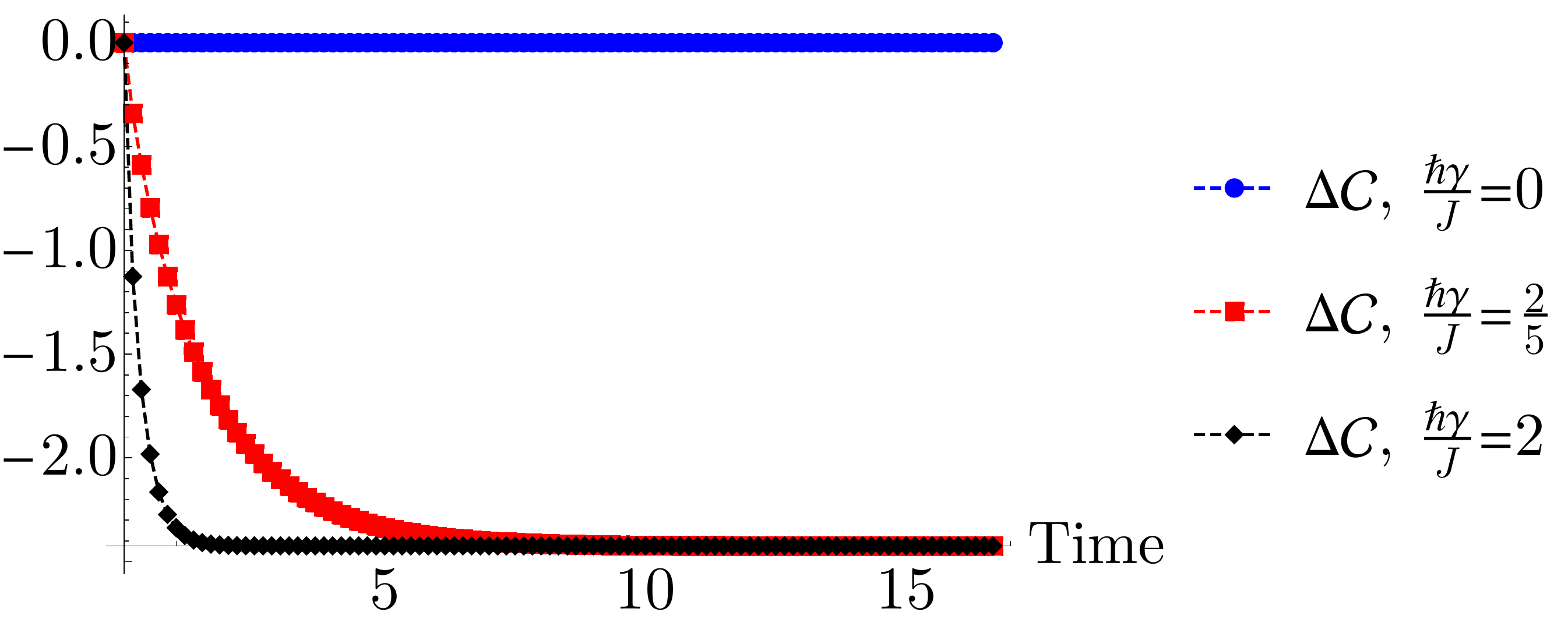}
\end{minipage}
\caption{Evolution of the change in mutual information ($\Delta \mathcal{I}$) and relative entropy of coherence ($\Delta \mathcal{C}$) for the MFI-model \eqref{ergodic} in its ergodic phase, $J=1$ and $g=1.05$, with $N=6$ and for the ``all-up'' state \eqref{eq:all-up}. Results were obtained as averages over $2 \times 10^2$.}
	\label{non}
\end{figure*}

The situation becomes more interesting for the mixed field Ising (MFI) model \cite{swingle,nie2019detecting,sahu2020information}. The MFI-model is closely related to the XXX-model. The major difference is that the nearest neighbor interactions are in the $z$-direction only, and an additional global field is added in the $x$-direction, see Fig.~\ref{mfi} for a sketch.
In formula we have
\begin{equation}
H_\mrm{MFI}=-J \sum_{i=1}^{N} \sigma_{i}^{z} \sigma_{i+1}^{z}- \sum_{i=1}^{N} h_{i}\sigma_{i}^{z}-g \sum_{i=1}^{N} \sigma_{i}^{x}.
\label{ergodic}
\end{equation}
In complete analogy to the XXX-model, we restrict ourselves to the nonintegrable parameter regime. To this end, we choose $J=1$, $g=1.05$, and $h_{i}$ are random variables drawn uniformly from $[-W,W]$ with $W=2$.  In fact, the magnitude of $W$ determines the speed with which information gets scrambled \cite{swingle}. The MFI-model exhibits scrambling for the ``all-up'' state \eqref{eq:all-up}. Figure~\ref{non} summarizes our numerical findings.

Interestingly, while the MFI-model does exhibit scrambling, even in the unitary case, $\gamma=0$, the model does not reach maximal scrambling, i.e., $\mathcal{I}(t)<2\ln(2)$ for all $t$. More strikingly, even the smallest amount of decoherence is sufficient to suppress all correlations, quantum as well as classical, in the system. After an initial increase the mutual information, $\mc{I}$ quickly drops to zero. This rather curious behavior can be understood by realizing that decoherence stands in direct competition to the last term of the Hamiltonian \eqref{ergodic}, which is the origin of scrambling in the $z$-direction. Thus, we immediately conclude that information scrambling is not generally robust against decoherence. Rather, it crucially depends on the competition of the terms in the Hamiltonian driving the scrambling dynamics and the interaction with $\mc{E}$. Otherwise, the MFI-model exhibits qualitatively similar behavior to what we found in the preceding models.

\paragraph*{Lipkin-Meshkov-Glick model.}

As a final example, we analyze the scrambling properties of the integrable Lipkin-Meshkov-Glick (LMG) model \cite{fogarty2020orthogonality,campbell2015shortcut,lipkin1965validity,ribeiro2007thermodynamical,ribeiro2008exact}. Its Hamiltonian reads,
\begin{equation}
H_{\mathrm{LMG}}=-\frac{J}{N} \sum_{i<j}^{N}\left(\sigma_{x}^{i} \sigma_{x}^{j}+\sigma_{y}^{i} \sigma_{y}^{j}\right)-\sum_{i=1}^{N} \sigma_{z}^{i}\,,
\label{lmgh}
\end{equation}
\begin{figure}
	\centering 
	\includegraphics[width=.46\textwidth]{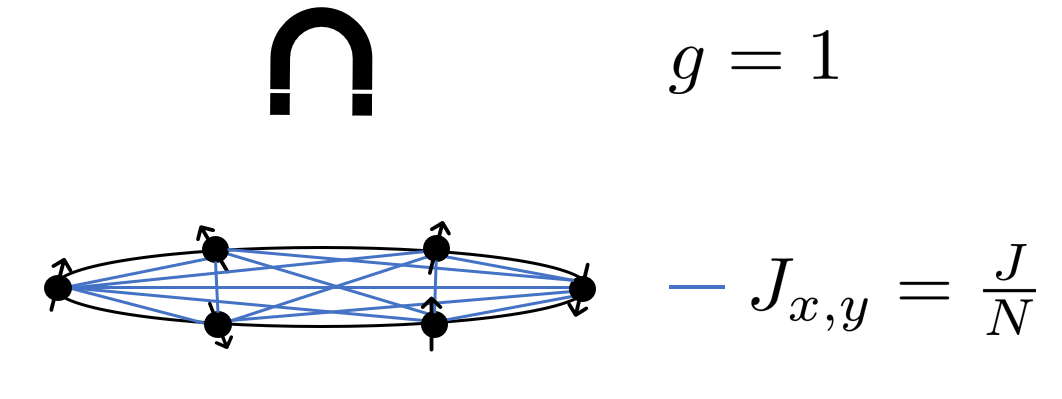}
	\caption{Sketch of the LMG-model with all-to-all interactions (in $x$ and $y$-directions), random local fields $h_i$, and a global field in $x$-direction.}
	\label{LMG}
\end{figure}
which is depicted in Fig.~\ref{LMG}.

The LMG-model is a critical spin system that undergoes a quantum phase transition at $J=1$. It is comprised of $N$ spin-1/2 particles with infinite-range interactions, under the action of a global transverse magnetic field.   The model was originally designed to study the shape phase transition in nuclei \cite{lipkin1965validity}, but its use was extended to other areas of physics, including shortcuts to adiabaticity and quantum speed limits \cite{fogarty2020orthogonality,campbell2015shortcut}. Interestingly, the LMG-model was also used to study the relationship between entanglement and quantum phase transitions \cite{lmg1,lmg2,lmg3,lmg4}.

\begin{figure*}
\begin{minipage}[l]{.48\textwidth}
Decoherence in computational basis\\
\vspace{.5em}
\includegraphics[width=\textwidth]{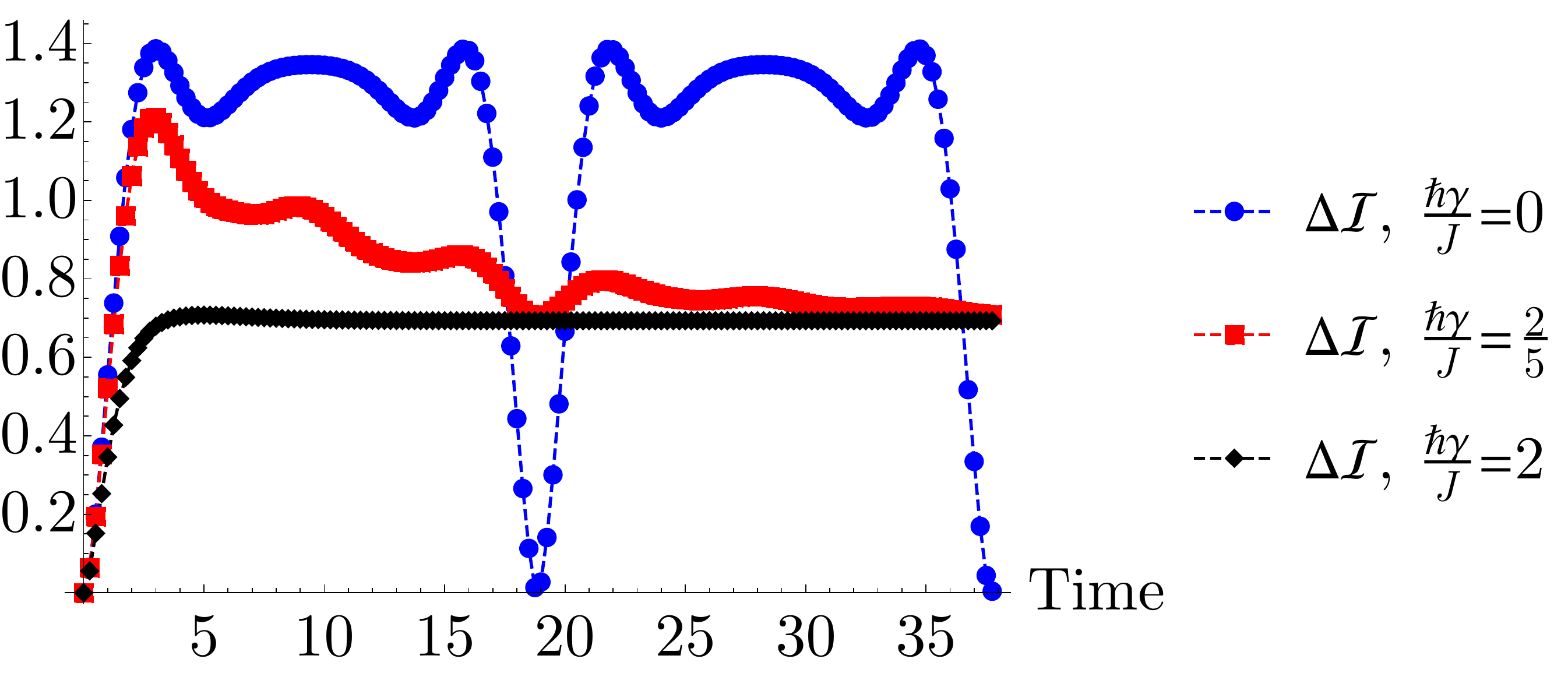}\\
\includegraphics[width=\textwidth]{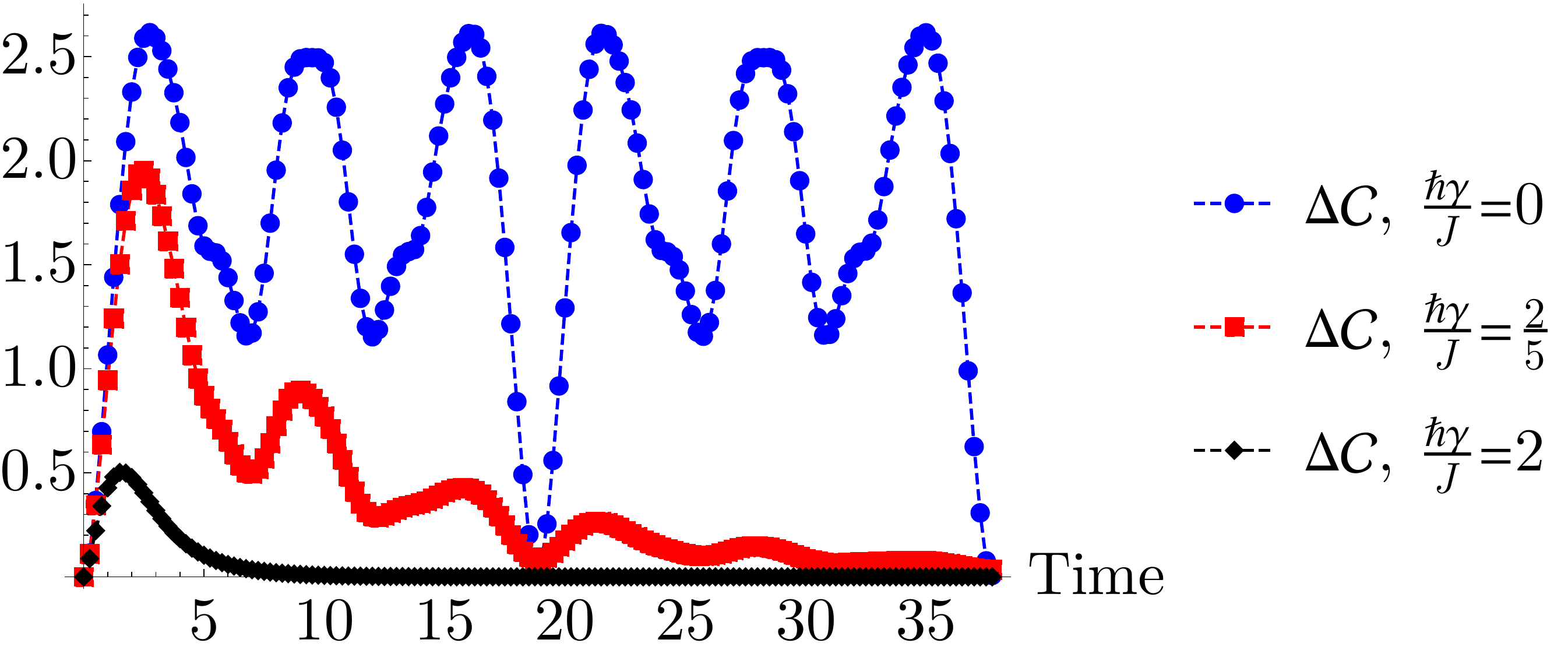}
\end{minipage}
\hfill
\begin{minipage}[l]{.48\textwidth}
Decoherence in energy basis\\
\vspace{.5em}
\includegraphics[width=\textwidth]{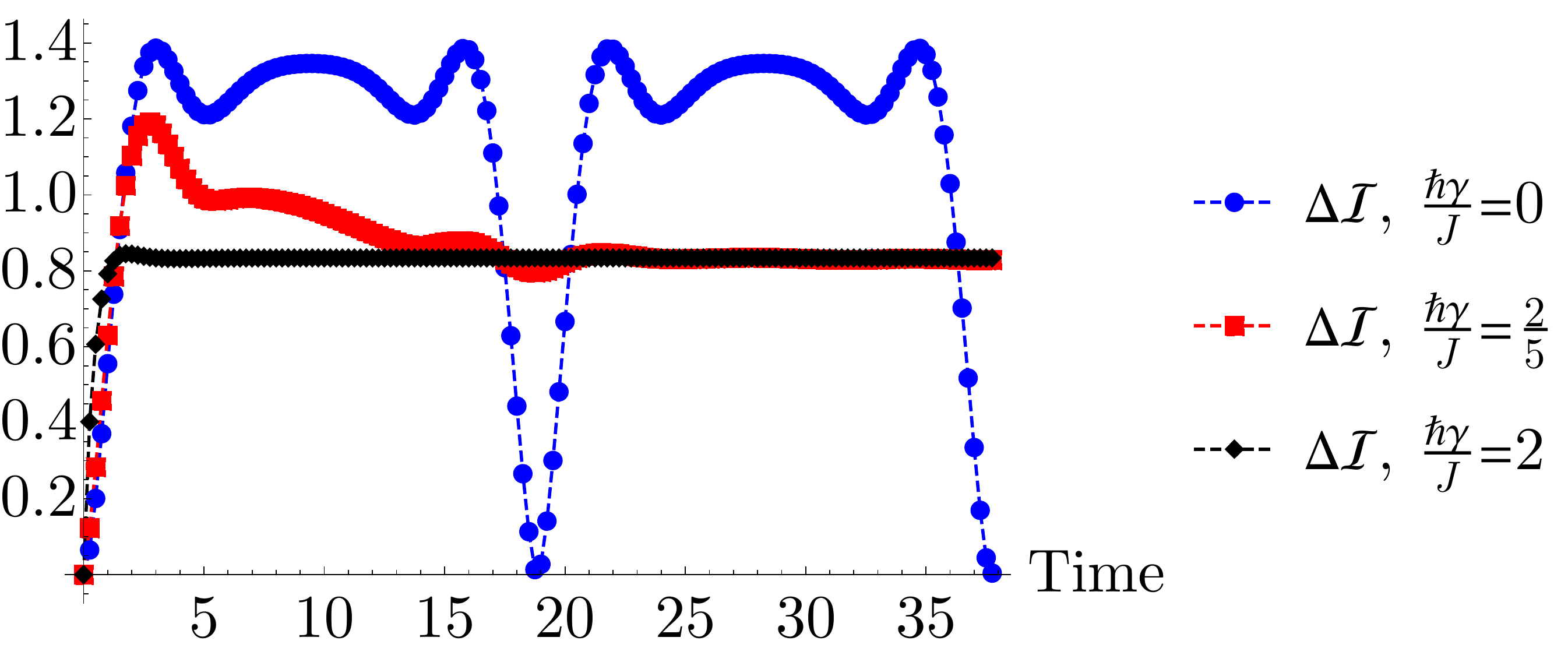}\\
\includegraphics[width=\textwidth]{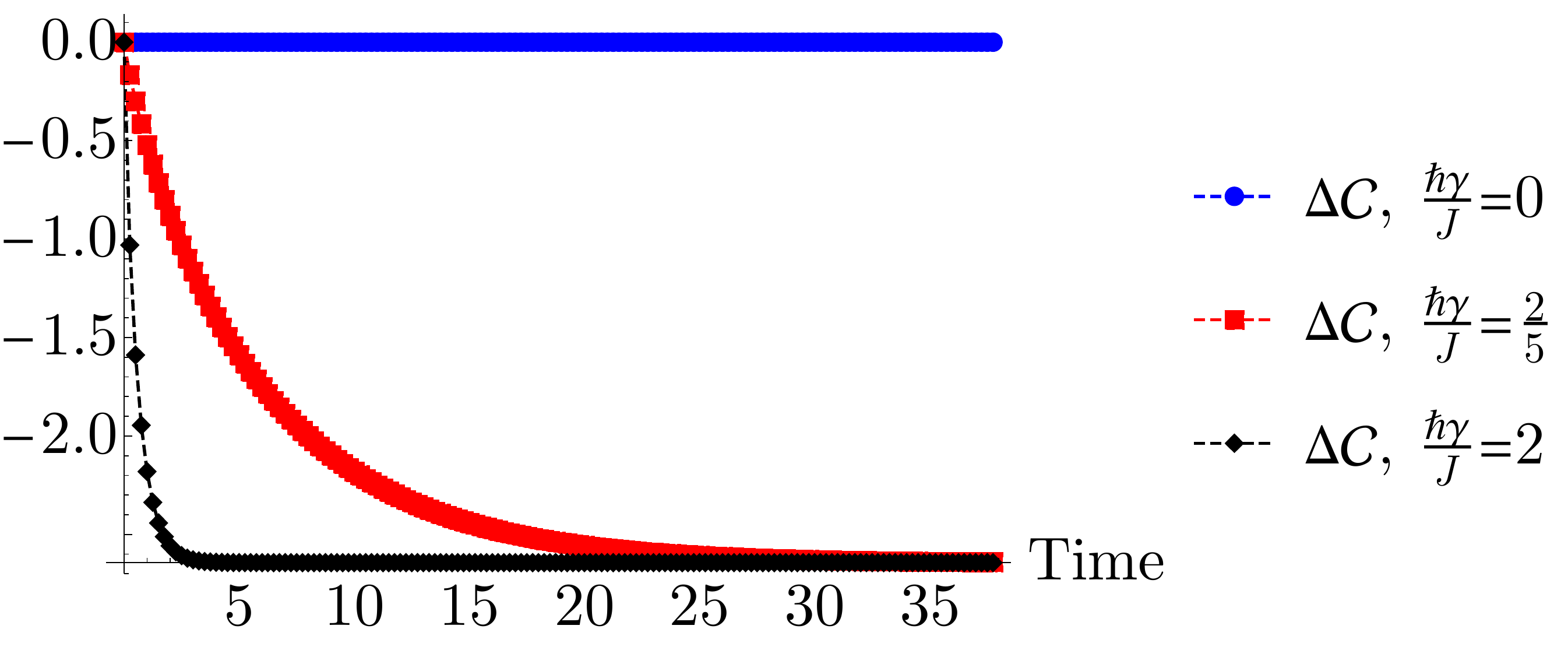}
\end{minipage}
	\caption{Evolution of the change in mutual information ($\Delta \mathcal{I}$) and relative entropy of coherence ($\Delta \mathcal{C}$) for the LMG-model \eqref{lmgh} with $N=6$ and an initial N\'eel state \eqref{eq:Neel}.}
	\label{lmg}
\end{figure*}

In contrast to the previous two spin chain models, the LMG-model is integrable in both phases, $J<1$ and $J>1$. Thus, one would not expect any fundamentally different behavior with respect to scrambling in each of the phases. Figure~\ref{lmg} collects our results for $J=1/2$ and an initial N\'eel state \eqref{eq:Neel}. 

Since the LMG-model is integrable, we do not see information scrambling in the strict sense for isolated dynamics. Rather, the model exhibits marked recurrences, which means that all information returns to the initial site. However, as soon as some decoherence is present, $\gamma>0$, the overall behavior becomes essentially indistinguishable to what we found for ergodic systems. This makes clear that in the presence of decoherence it is a rather involved task to identify whether a system exhibits genuine information scrambling, or whether information is simply exchanged between $\mc{S}$ and the environment. Therefore, without prior knowledge about whether or not a system exhibits scrambling, verifying information scrambling in realistic settings appears challenging. Neither the OTOC nor the mutual information give a clear indication on whether information is ``dumped'' into the environment or scrambled with the system itself.

\section{Concluding remarks}

In the present work we analyzed the effects of decoherence on information scrambling at the conceptual level and through numerical case studies. For the conceptual framework, we derived a general fluctuation theorem for open quantum systems, specifically for stochastic variables linked to the mutual information between system and environment. We related the competing effects of scrambling and decoherence to their respective contributions to the entropy production. Remarkably, choosing the mutual information as a quantifier clearly shows the additive contributions that affect the flow of quantum information in open quantum systems.

In the numerical part of our analysis, we studied the scrambling dynamics of $\mathcal{S}$ under decoherence in five unique models, showing the behavior of the mutual information and relative entropy of coherence in each model. At least qualitatively, we found universal behavior. Any deviation from the monotonic growth of the mutual information and relative entropy of coherence points to some outside interaction with the environment, given that the unitary dynamics are indeed scrambling. This interaction might be in the form of pure decoherence (destruction of coherences + no dissipation), or destruction of coherences accompanied with dissipation. Notably, the behavior of the chosen quantifiers ($\mathcal{I}$ and $\mathcal{C}$) is similar across the different models, with the exception of some very specific scenarios seen in condensed matter spin chain models (such as full recurrences in case of integrable dynamics). Therefore, we are reasonably confident that our findings also apply to quantum gravity models to be studied in future experiments \cite{chew2017approximating,chen2018quantum,marino2019cavity,lewis2019unifying,alavirad2019scrambling,bentsen2019treelike,yin2020bound}.

The conceptual notions and the gained insight of our work may open the door for further inquiry, such as the study of quantum to classical transitions in the context of information scrambling. To study such transitions one would need to go beyond describing decoherence at the level of $\mc{S}$ and have full access to the degrees of freedom of $\mathcal{E}$. Our fluctuation theorem seems to be uniquely suited to be generalized to also include quantum discord, which can be written as the difference between two different measures of mutual information \cite{ollivier2001quantum}.  Quantum discord is a measure of the ``quantumness'' of the correlations between $\mathcal{S}$ and $\mathcal{E}$. Therefore, quantum discord might represent a valid quantifier of scrambling in open quantum systems, and a good starting point to understand quantum to classical transitions from an information scrambling perspective.  However, actually computing quantum discord is a challenge on its own, which is why we leave this analysis for future work.

\acknowledgements{This research was supported by grant number FQXi-RFP-1808 from the Foundational Questions Institute and Fetzer Franklin Fund, a donor advised fund of Silicon Valley Community Foundation (SD).}

\appendix

\section{Master equation}
\label{a}

Using the notation introduced in Ref.~\cite{touil2020quantum}, \myeqref{master1} can be simplified to give
\begin{equation}
\frac{\pd |\rho\rangle}{\pd t}=\left(W-\gamma\,\mbb{I}_{N^{2}}+\gamma V\right)|\rho\rangle\equiv A|\rho\rangle\,\,.
\label{fock1}
\end{equation}
Here, $W=-i/\hbar\,\left(H \otimes \mbb{I}_{N}-\mbb{I}_{N} \otimes H^{\top}\right)$, and $\mbb{I}_{N^{2}}$ and $\mbb{I}_{N}$ are identity matrices of dimensions $N^{2} \times N^{2}$ and $N \times N$ respectively. Finally,
\begin{equation}
V=\sum_{\mathcal{K}=1}^{N}|(\mathcal{K}-1)N+\mathcal{K}\rangle\langle (\mathcal{K}-1)N+\mathcal{K}|,
\end{equation}
where the vectors $|\mathcal{L}\rangle$ refer to $N^{2}$-dimensional vectors that live in Fock-Liouville space, such that all entries are zero except the $\mathcal{L}$th entry with a value of one. These vectors are not to be confused with the basis vectors where we define decoherence, as they live on two different Hilbert spaces.

Now that we defined the matrices in \myeqref{fock1} we mapped our problem to a dynamical equation with a linear vector field described by the operator/matrix $A$ in Fock-Liouville space. Also, \myeqref{master1} is equivalent to a dynamical CPTP map of the evolution of our density matrix (i.e. taking a density matrix to another density matrix in Hilbert space). Therefore the maximal forward interval of existence of the solutions of \myeqref{fock1} is $[0, +\infty[$. In other words, our solutions $|\rho(t)\rangle$ are well-defined (in terms of existence uniqueness),
\begin{equation}
\left(\forall t \in [0, +\infty[\right); \  |\rho(t)\rangle=\e{A\,t}\,|\rho(0)\rangle.
\label{fock2}
\end{equation}
Equation~\eqref{fock2} is the general solution of the master equation in Fock-Liouville space. The solutions give a density vector that can be directly mapped to a density matrix in Hilbert space.

\bibliography{deco}

\end{document}